%% file: main_qqtth.tex
\documentclass[11pt,a4paper]{article}
\pdfoutput=1

\usepackage{jheppub}
\usepackage{graphicx,epsfig,amsmath,amssymb}
\usepackage{prettyref}
\usepackage{subcaption}
\usepackage{enumitem}
\usepackage{calc}
\usepackage{mathtools}
\usepackage{float}
\usepackage{booktabs}

\input{all.tikzdefs}
\input{all.tikzstyles}

\usepackage{hyperref}
\hypersetup{pdfusetitle,colorlinks=true,urlcolor=purple,citecolor=purple,linkcolor=purple}

\let\origref\ref
\newrefformat{chap}{\hyperref[#1]{Chapter~\origref*{#1}}}
\newrefformat{app}{\hyperref[#1]{Appendix~\origref*{#1}}}
\newrefformat{sec}{\hyperref[#1]{Section~\origref*{#1}}}
\newrefformat{subsec}{\hyperref[#1]{Section~\origref*{#1}}}
\newrefformat{tab}{\hyperref[#1]{Table~\origref*{#1}}}
\newrefformat{fig}{\hyperref[#1]{Figure~\origref*{#1}}}
\newrefformat{eq}{eq.\,\hyperref[#1]{(\origref*{#1})}}
\AtBeginDocument{\renewcommand*{\ref}[1]{\prettyref{#1}}}

\graphicspath{{./plots}}

\def\nn{\nonumber}
\def\as{\alpha_s}
\def\aspi{\frac{\alpha_s}{2\pi}}
\def\asb{\alpha_s^0}
\def\asl{\alpha_s^{(n_l)}}
\def\aslpi{\frac{\alpha_s^{(n_l)}}{2\pi}}

\def\eps{\varepsilon}

\def\be{\begin{equation}}
\def\ee{\end{equation}}
\def\bea{\begin{eqnarray}}
\def\eea{\end{eqnarray}}
\def\ket#1{|{#1}\rangle}
\def\bra#1{\langle{#1}|}
\def\braket#1#2{\langle #1 |#2 \rangle}

\def\lmu{l_\mu}
\def\d{\mathrm{d}}
\newcommand{\secdec}{\textsc{SecDec}{}}
\newcommand{\pysecdec}{py\secdec}
\newcommand{\fstt}{\mathrm{frac}_{s_{t\bar{t}}}}
\newcommand{\gosam}{\textsc{GoSam}{}}

\newcommand{\mt}{m_t}

\newcommand{\MSbar}{\overline{\text{MS}}}
\title{Two-loop amplitudes for $t\bar{t}H$ production: the quark-initiated $N_f$-part}
\hypersetup{pdftitle={ttH production at two loops: the quark-initiated Nf-part}}
\author[a]{Bakul Agarwal,}
\author[a]{Gudrun Heinrich,}
\author[b]{Stephen P.~Jones,}
\author[a]{Matthias Kerner,}
\author[c]{Sven Yannick Klein,}
\author[a]{Jannis Lang,}
\author[a]{Vitaly Magerya,}
\author[a]{Anton Olsson}

\affiliation[a]{Institute for Theoretical Physics, Karlsruhe Institute of Technology, Wolfgang-Gaede-Str.~1, 76131 Karlsruhe, Germany}
\affiliation[b]{Institute for Particle Physics Phenomenology, Durham University, South Road, Durham DH1 3LE, UK}
\affiliation[c]{Institute for Theoretical Particle Physics and Cosmology, RWTH Aachen University, Sommerfeldstr.~16, 52074 Aachen, Germany}

\emailAdd{bakul.agarwal@kit.edu}
\emailAdd{gudrun.heinrich@kit.edu}
\emailAdd{stephen.jones@durham.ac.uk}
\emailAdd{matthias.kerner@kit.edu}
\emailAdd{jannis.lang@kit.edu}
\emailAdd{sven.yannick.klein@rwth-aachen.de}
\emailAdd{vitalii.maheria@kit.edu}
\emailAdd{anton.olsson@kit.edu}

\preprint{{\small  IPPP/24/03, KA-TP-02-2024, P3H-24-007, TTK-24-03}}

\abstract{
  We present numerical results for the two-loop virtual amplitude entering the
  NNLO corrections to Higgs boson production in association with a top quark pair at the LHC,
  focusing, as a proof of concept of our method, on the part of the quark-initiated channel containing loops of massless or massive quarks. 
  Results for the UV renormalised and IR subtracted two-loop amplitude for each colour structure are given at selected phase-space points and visualised in terms of surfaces as a function of two-dimensional slices of the full phase space.
}

\begin{document}

\maketitle
\newpage

\section{Introduction}

\input{intro.tex}

\section{Description of the method}
\label{sec:method}

\input{qq_method.tex}

\section{Results}
\label{sec:results}

\input{results.tex}

%\clearpage

\section{Conclusions}
\label{sec:conclusions}

We have presented numerical results entering $t\bar{t}H$ production at NNLO QCD, for the quark initiated $N_f$--parts of the two-loop amplitude including loops of both massless and massive quarks.
This calculation serves as a proof of concept that our setup is capable of calculating two-loop pentagon amplitudes with internal massive propagators and three massive particles in the final state.
We have performed the UV renormalisation and subtraction of IR poles, presenting the finite part of the two-loop amplitude, split into nine different colour structures for a general colour group.

For the reduction to master integrals, we do not attempt to obtain a fully symbolic reduction and instead perform a numerical reduction for each phase-space point leaving the dimensional regulator symbolic.
The master integrals are evaluated with a recent version of {\pysecdec}, which has been further extended to support integration over double-double precision integrands, this allows us to obtain stable results also in the high-energy and collinear limits where many digits of the master integrals cancel.
The evaluation times vary substantially over the phase space, being of
the order of five minutes in the bulk of the phase space, increasing substantially when
approaching the $\beta\to 1$ limit.
We do not expect the full quark channel to present further major obstacles within our calculational framework.

Although we have demonstrated that the amplitude can be evaluated with sufficient precision at individual phase-space points, the largest remaining challenge for producing realistic phenomenological applications is to sufficiently densely sample the full 5-dimensional phase-space.
One possible way of addressing this obstacle is to supplement the evaluated phase-space points with a reliable interpolation framework that allows data points at any 5-dimensional phase-space point to be provided with sufficient accuracy.
This is a challenge for kinematic regions where the amplitude has a very steep gradient, for example in the high-energy region with quasi-collinear configurations. While an interpolation covering the whole phase space is feasible, assessing the associated uncertainties is challenging; this is work in progress.

\section*{Acknowledgements}

This research was supported by  the  Deutsche  Forschungsgemeinschaft
(DFG, German Research Foundation) under grant 396021762 - TRR 257. 
SJ is supported by a Royal Society University Research Fellowship (Grant URF/R1/201268) 
and by the UK Science and Technology Facilities Council under contract ST/T001011/1.

\clearpage

\appendix

% \section{Appendix}
\section{Additional details}
\label{app:appendix}
\input{appendix.tex}

\bibliographystyle{JHEP}

\bibliography{main_qqtth}

\end{document}

%% file: all.tikzstyles
\tikzstyle{blank}=[fill=white, shape=circle, draw=white, inner sep=0.8pt]
\tikzstyle{dot}=[fill=black, shape=circle, draw=black, inner sep=0.8pt]
\tikzstyle{fat}=[fill=white, shape=circle, draw={rgb,255: red,176; green,36; blue,39}, dashed, line width=1pt, inner sep=0.8pt]

\tikzstyle{edge}=[-, draw=EdgeColor, line width=1.2pt, line cap=rect, style=whitebg]
\tikzstyle{incoming edge}=[|-, style=edge, draw=PaleEdgeColor, style=arrowin]
\tikzstyle{outgoing edge}=[->, style=edge, draw=PaleEdgeColor, style=arrow]

\tikzstyle{massive edge}=[-, draw=MassiveEdgeColor, line width=2pt, style=whitebg, line cap=rect]
\tikzstyle{incoming massive edge}=[|-, style=massive edge, draw=PaleMassiveEdgeColor, style=arrowin]
\tikzstyle{outgoing massive edge}=[->, style=massive edge, draw=PaleMassiveEdgeColor, style=arrow]

\tikzstyle{fermion}=[-, draw=FermionColor, line width=1pt, style=whitebg, line cap=rect, postaction=decorate, decoration={markings,mark=at position .60 with {\arrow{stealth[round]}}}]
\tikzstyle{incoming fermion}=[-, style=fermion, draw=PaleEdgeColor]
\tikzstyle{outgoing fermion}=[-, style=fermion, draw=PaleEdgeColor]

\tikzstyle{massive fermion}=[-, draw=MassiveFermionColor, line width=1.9pt, line cap=rect, style=whitebg]
\tikzstyle{incoming massive fermion}=[-, style=massive fermion, draw=PaleEdgeColor]
\tikzstyle{outgoing massive fermion}=[-, style=massive fermion, draw=PaleEdgeColor, style=arrow]

\tikzstyle{gluon}=[-, draw=GluonColor, line width=1pt, preaction={{draw=white,line width=2pt}}, line cap=rect, decorate, decoration={coil,aspect=1.5,segment length=7pt}]
\tikzstyle{incoming gluon}=[-, style=gluon, draw=PaleEdgeColor]
\tikzstyle{outgoing gluon}=[-, style=gluon, draw=PaleEdgeColor]

\tikzstyle{ghost}=[-, style=fermion, line width=1pt, line cap=round, dash pattern={on 0pt off 3\pgflinewidth}]
\tikzstyle{incoming ghost}=[-, style=ghost, draw=PaleEdgeColor]
\tikzstyle{outgoing ghost}=[-, style=ghost, draw=PaleEdgeColor]

\tikzstyle{scalar}=[-, draw=ScalarColor, line width=1pt, densely dashed]
\tikzstyle{incoming scalar}=[-, style=scalar, draw=PaleScalarColor]
\tikzstyle{outgoing scalar}=[-, style=scalar, draw=PaleScalarColor]

\tikzstyle{massive scalar}=[-, draw=ScalarColor, line width=2pt, densely dashed]
\tikzstyle{incoming massive scalar}=[-, style=massive scalar, draw=PaleScalarColor]
\tikzstyle{outgoing massive scalar}=[->, style=massive scalar, draw=PaleScalarColor, style=arrow]

\tikzstyle{vector}=[-, draw=VectorColor, line width=1pt, preaction={{draw=white,line width=2pt}}, line cap=rect, decorate, decoration=snake]
\tikzstyle{incoming vector}=[-, style=vector, draw=PaleEdgeColor]
\tikzstyle{outgoing vector}=[-, style=vector, draw=PaleEdgeColor]

\tikzstyle{dot1}=[-, draw=none, postaction=decorate, decoration={markings,mark=at position .50 with {\node[style=dot]{};}}]
\tikzstyle{dot2}=[-, draw=none, postaction=decorate, decoration={markings,mark=between positions 0.33 and 0.67 step 0.33 with {\node[style=dot]{};}}]
\tikzstyle{dot3}=[-, draw=none, postaction=decorate, decoration={markings,mark=between positions 0.25 and 0.76 step 0.25 with {\node[style=dot]{};}}]
\tikzstyle{dot4}=[-, draw=none, postaction=decorate, decoration={markings,mark=between positions 0.20 and 0.81 step 0.20 with {\node[style=dot]{};}}]

%% file: intro.tex
Higgs production in association with a top quark pair was observed for the first time a few years ago at the Large Hadron Collider (LHC)~\cite{Aaboud:2018urx,Sirunyan:2018hoz,CMS:2018fdh} and will play an important role at the High-Luminosity (HL) LHC.
The process $pp\to t\bar{t}H$ is particularly interesting due to its direct sensitivity to the top-Yukawa coupling $y_t$, which is now being constrained with increasing accuracy, including potential CP-violating couplings~\cite{CMS:2022dbt,ATLAS:2023cbt}.
The importance of this process was realised a long time ago~\cite{Kunszt:1984ri,Ng:1983jm}, and NLO QCD corrections for on-shell $t\bar{t}H$ production have been known for many
years~\cite{Beenakker:2001rj,Reina:2001sf,Beenakker:2002nc,Dawson:2002tg,Dawson:2003zu}.
The corrections have been matched to parton showers in Refs.~\cite{Frederix:2011zi,Garzelli:2011vp,Hartanto:2015uka}.
NLO EW corrections have first been calculated in Ref.~\cite{Frixione:2014qaa}, the EW corrections have been 
combined with NLO QCD  corrections within the narrow-width-approximation (NWA) for top-quark decays in Refs.~\cite{Yu:2014cka,Frixione:2015zaa}.
NLO QCD corrections to off-shell top quarks in $t\bar{t}H$ production with leptonic $W$-decays have been calculated in Ref.~\cite{Denner:2015yca,Stremmer:2021bnk} and full off-shell effects with $H\to b\bar{b}$ have been calculated in Refs.~\cite{Denner:2020orv,Bevilacqua:2022twl}.
A combination of the NLO QCD corrections with NLO EW corrections has been presented in Ref.~\cite{Denner:2016wet}, NLO QCD corrections combined with electroweak Sudakov logarithms and a parton shower have been studied in Ref.~\cite{Pagani:2023wgc}.

Soft gluon resummation at NLO+NNLL has been performed in
Refs.~\cite{Kulesza:2015vda,Broggio:2015lya,Broggio:2016lfj,Kulesza:2017ukk}, see also Ref.~\cite{vanBeekveld:2020cat}, soft and Coulomb corrections have been resummed in Ref.~\cite{Ju:2019lwp}.
The NLO+NNLL  resummed results have been further improved  including also the processes $t\bar{t}W^\pm, t\bar{t}Z$~\cite{Kulesza:2018tqz,Broggio:2019ewu}, where Ref.~\cite{Broggio:2019ewu} also includes EW corrections.

Given  the projection that the statistical uncertainty will shrink to the order of 2-3\% after 3000 fb$^{-1}$~\cite{Azzi:2019yne},
the measurement of $t\bar{t}H$ will be dominated by systematics.
As the dominant systematic uncertainties currently come from modelling uncertainties of signal and backgrounds~\cite{Aaboud:2018urx,Sirunyan:2018hoz,CMS:2018fdh}, there is a clear need to reduce the theory uncertainties.
At NLO QCD the scale uncertainties are of the order of 10-15\%, therefore NNLO QCD corrections are necessary to match the experimental precision at the HL--LHC.

First steps towards this goal already are available in the literature: in Ref.~\cite{Catani:2021cbl}, ${\cal O}(\alpha_s^4)$ corrections to the flavour-off-diagonal channels have been calculated, exploiting relations from $q_T$-resummation~\cite{Catani:2014qha}.
In Ref.~\cite{Catani:2022mfv}, the total NNLO cross section has been presented,  where  for the finite part of the two-loop virtual amplitude  a soft Higgs boson approximation has been used.
The coefficients of the two-loop infrared singularities for this process have been calculated in Ref.~\cite{Chen:2022nxt}.
In Ref.~\cite{Brancaccio:2021gcz}, the order ${\cal O}(y_t^2\alpha_s)$ corrections to the perturbative fragmentation functions and to the splitting functions relevant for associated top-Higgs production have been calculated.
Analytic results for the master integrals entering the leading-colour two-loop 
amplitudes that are proportional to the number of light flavours for the processes  $gg, q\bar{q}\to t\bar{t}H$ have recently been presented in Ref.~\cite{FebresCordero:2023gjh}.
Furthermore, the $gg \to t\bar{t}H$ one-loop amplitude has been calculated semi-numerically up to second order in the $\eps$-expansion in Ref.~\cite{Buccioni:2023okz}.
Results for the two-loop amplitudes for both the gluon and the quark channel in the high-energy boosted limit have been provided very recently in Ref.~\cite{Wang:2024pmv}.

In this work we present numerical results for the two-loop
virtual amplitudes for $q\bar{q}\to t\bar{t}H$ which contain
closed fermion loops, i.e.\ are proportional to the number of
light fermion flavours $n_l$, heavy fermion flavours $n_h$, or
both.
Specifically, we calculate the renormalised interference of
the two-loop amplitude with the tree-level amplitude, with
full dependence on the top quark and Higgs masses, split into nine
independent colour and fermion flavour factors.
Many of the master integrals appearing in this calculation are 
not currently known fully analytically, we therefore choose
to evaluate all integrals using the sector decomposition~\cite{Hepp:1966eg,Roth:1996pd,Binoth:2000ps,Heinrich:2008si} approach.
Our results are visualised on one- and two-dimensional slices of the five-dimensional phase space.
These results can be regarded as a proof of concept for the calculation of the other colour structures and partonic channels.

The paper is structured as follows.
In~\ref{sec:method} we describe the kinematics of the process, the structure of the $q\bar{q}\to t\bar{t}H$ amplitude, and the workflow of our calculation.
We present our results in~\ref{sec:results} and conclude in~\ref{sec:conclusions}.
Further details of our calculation, including the UV renormalisation, the colour decomposition, the integral families used for the integral reduction, and full numeric results at several example phase-space points are given in~\ref{app:appendix}.

%% file: qq_method.tex
The calculation of the virtual two-loop amplitudes contains the channels $q\bar{q}\to t\bar{t}H$ and $gg\to t\bar{t}H$.
Here we focus on the quark initial state.

\subsection{Kinematics}

We use ``all incoming'' kinematics,
\be
q(p_1)+\bar{q}(p_2)\to t(-p_3)+\bar{t}(-p_4)+H(-p_5)\;,
\ee
such that 
\begin{equation}
    p_1^2 = p_2^2 = 0, \qquad p_3^2=p_4^2=m_t^2, \qquad p_5^2 = m_H^2,
\end{equation}
and the Mandelstam invariants are defined by
\begin{equation}
    s_{ij} = (p_i+p_j)^2.
\end{equation}
Ten such invariants can be built; five of them are independent due to momentum conservation.
Out of these we use the following dimensionless variables:
\begin{equation}
  x_{12}=\frac{s_{12}}{m_t^2}, \;
  x_{23}=\frac{s_{23}-m_t^2}{m_t^2}, \;
  x_{35}=\frac{s_{35}-m_t^2}{m_t^2}, \;
  x_{45}=\frac{s_{45}-m_t^2}{m_t^2}, \;
  x_{14}=\frac{s_{14}-m_t^2}{m_t^2}.
  \label{eq:variable_def}
\end{equation}
There is also an independent parity odd invariant,
\begin{align}
\epsilon(1234) = 4 i \epsilon_{\mu \nu \rho \sigma } p_1^{\mu} p_2^{\nu} p_3^{\rho} p_4^{\sigma} = {\rm tr}( \gamma_{5} \not{p}_{1}  \not{p}_{2}  \not{p}_{3} \not{p}_{4}) \,.
\end{align}
The square of the parity odd invariant is equal to the Gram determinant spanned by four linearly independent external momenta and is not algebraically independent of the other invariants.
However, as $\epsilon(1234)$ picks up a sign under parity, while the square root of the Gram determinant does not, the sign of the parity odd invariant must be specified to fully describe a physical phase-space point.
QCD is invariant under parity, therefore, the QCD corrections to the $t\bar{t}H$ production amplitudes ultimately must not depend on the sign of the invariant.

\subsection{Phase space parametrisation}

\begin{figure}
    \centering
    \input{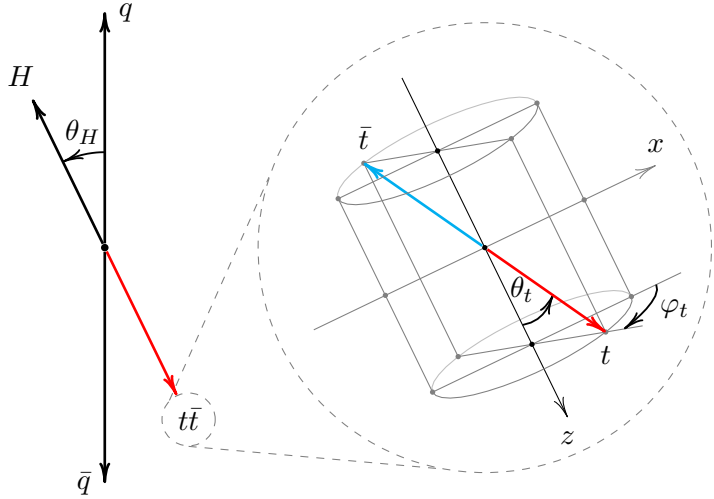}
    \caption{%
        The phase-space parameters.
        The angles $\theta_t$ and $\varphi_t$ are local to the
        $t\bar{t}$ rest frame, while $\theta_H$ is local to the
        $t\bar{t}H$ rest frame.
    }
    \label{fig:PS}
\end{figure}

The phase-space volume for $t\bar{t}H$ production is non-trivial
when expressed in the variables given in \ref{eq:variable_def}.
To parametrise it in a more explicit way we factorise it into sub-phase-space volumes
for the production of a ``$t\bar{t}$ state'' and a Higgs boson, combined with the ``decay'' of the $t\bar{t}$ state into two top quarks, leading to the following expression:
\begin{align}
    \label{eq:phase_factor}
    \d{}\Phi_{t\bar{t}H}
    &=\frac{1}{2^{10} \pi^4 \hat{s}\,s_{t\bar{t}}}\sqrt{\lambda(s_{t\bar{t}},m_t^2,m_t^2)}\sqrt{\lambda(\hat{s},s_{t\bar{t}},m_H^2)}\times\nonumber \\
    &\quad \Theta(\sqrt{\hat{s}}-2m_t-m_H)\Theta(s_{t\bar{t}}-4m^2_t)\Theta([\sqrt{\hat{s}}-m_H]^2-s_{t\bar{t}})\, \d{} s_{t\bar{t}}\, \d{}\Omega_{t\bar{t}} \, \mathrm{sin}\theta_H\, \d{} \theta_H
\end{align}
with $\d{} \Omega_{t\bar{t}} = \mathrm{sin}\theta_t \, \d{} \theta_t \, \d{} \varphi_t$, the K\"all\'en function $\lambda(a,b,c)=a^2+b^2+c^2-2ab-2bc-2ca$,
\begin{align}
    \label{eq:sstt_def}
    &\hat{s} = (p_3 + p_4 + p_5)^2,&
    &s_{t\bar{t}} = (p_3 + p_4)^2,&
\end{align}
and the angles $\theta_H$, $\theta_t$, and $\varphi_t$ introduced as in \ref{fig:PS} with precise definitions given in \ref{app:PS}.

As the production threshold of the $t\bar{t}H$ system is located at $s_0=(2m_t+m_H)^2$, a convenient
variable for a scan in partonic energy is
\begin{equation}
    \beta^2 = 1-\frac{s_0}{\hat{s}}
    \label{eq:beta_def}
\end{equation}
such that $\beta^2\to 0$ at the production threshold and $\beta^2\to 1$ in
the high energy limit.

For a compact parametrisation of the fraction of kinetic energy which
enters the $t\bar{t}$ system, we define the variable $\fstt$ as
\begin{align}
    \fstt = \frac{s_{t\bar{t}}-4m_t^2}{\left(\sqrt{\hat{s}}-m_H\right)^2-4m_t^2}\;,
    \label{eq:fstt_def}
\end{align}
with $\fstt=0$ corresponding to the production threshold of the
$t\bar{t}$ system with the Higgs boson carrying the remaining energy, and $\fstt=1$ corresponding to the production
threshold of the Higgs boson, with the $t\bar{t}$ system carrying the remaining energy.
Note that if the phase-space integration is performed in $\fstt$,
a Jacobian factor of ${\d{} s_{t\bar{t}}}/{\d{} \fstt}$ has
to be included for the full phase-space density of
\begin{equation}
    \frac{ \d{}\Phi_{t\bar{t}H} }{ \d{} \fstt \, \d{} \theta_H \d{} \theta_t \, \d{} \varphi_t} = 
    \frac{ \sqrt{ \vphantom{m_H^2} \lambda(s_{t\bar{t}},m_t^2,m_t^2)} \sqrt{\lambda(\hat{s},s_{t\bar{t}},m_H^2)} }{2^{10} \pi^4 \hat{s}\,s_{t\bar{t}}}
    \left( \left(\sqrt{\hat{s}}-m_H\right)^2-4m_t^2 \right)
    \mathrm{sin}\theta_H \,
    \mathrm{sin}\theta_t .
    \label{eq:PSdensity}
\end{equation}
The set of parameters $\{\beta^2, \fstt, \theta_H, \theta_t,
\varphi_t\}$ provide a way to parametrise the amplitude
which is equivalent to using the five invariants from \ref{eq:variable_def}; the mapping
between them is defined by \ref{eq:sstt_def}, \ref{eq:beta_def},
\ref{eq:fstt_def}, and the relations given in \ref{app:PS}.
In these parameters the physical region of the phase space is found as
\begin{equation}
    \beta^2\in [0, 1],\ \mathrm{frac}_{s_{t\bar{t}}}\in [0, 1],\ \theta_H\in [0, \pi],\ \theta_t\in [0, \pi],\ \varphi_t\in [0, 2\pi].
    \label{eq:parameters}
\end{equation}
Note that the probability density of \ref{eq:PSdensity} will suppress the low-$\beta$
region as $\beta^4$, and enhance the high-$\beta$ region as ${1/(1-\beta)^2}$.
It will also suppress both low- and high-$\fstt$ regions as $\sqrt{\fstt}$ and $\sqrt{1-\fstt}$, respectively.
Nominally, the factors $\mathrm{sin}\theta_H$ and $\mathrm{sin}\theta_t$ also suppress the polar cap regions in $\theta_H$ and $\theta_t$, but this is only an artifact of the choice to map the respective spherical regions to a hypercube.

\subsection{Amplitude symmetries}
\label{sec:symmetries}

The squared amplitude of $q\bar{q}\to t\bar{t}H$ is invariant under the following transformations:
\begin{itemize}
    \item
        $\varphi_t \to -\varphi_t$, due to parity invariance;
    \item
        $\{\theta_H, \theta_t\} \to \{\theta_H+\pi, \theta_t+\pi\}$,
        which is the simultaneous swapping of $q$ with $\bar{q}$, and $t$ with $\bar{t}$;
    \item
        $\theta_H \to \theta_H \pm 2\pi$, and $\{\theta_H, \theta_t\} \to \{-\theta_H, -\theta_t\}$ due to $\theta_H$ being a polar angle;
    \item
        $\theta_t \to \theta_t \pm 2\pi$, $\varphi_t \to \varphi_t \pm 2\pi$, and $\{\theta_t, \varphi_t\} \to \{-\theta_t, \varphi_t + \pi\}$ due to $\theta_t$ and $\varphi_t$ being polar coordinates on a sphere;
    \item
        $\varphi_t \to \text{anything}$ when $\theta_t=0$ or $\theta_t=\pi$ due to analyticity of the amplitude at the poles of the sphere in $(\theta_t, \varphi_t)$.
\end{itemize}

\subsection{Ultraviolet renormalisation}
\label{sec:uv-renormalisation}

To produce an ultraviolet (UV) and infrared (IR) finite 2-loop
amplitude, we work in conventional dimensional regularisation
assuming that all momenta live in $d=4-2\eps$ space-time dimensions,
and use the expressions for the two-loop singularity structure of
massive amplitudes worked out in Ref.~\cite{Ferroglia:2009ii},
also used in Ref.~\cite{Baernreuther:2013caa}.

We expand the bare amplitude for $q\bar{q}\to t\bar{t}H$ in the strong coupling as
%\footnote{The explicit masses in the arguments of the amplitudes denote the parameters of the Lagrangian, i.e.\ they only enter in the propagators.  Kinematic parameters such as $s_{ij}$ and $p_i^2$ are not explicitly listed in the description of the renormalisation procedure.}
%
\begin{align}
   {\cal A}^b(\asb,y_t^0,m^0,\eps) 
  & = 
  4 \pi \asb \frac{y_t^0}{\sqrt{2}} \left[
   {\cal A}_0^{b}(m^0,\eps) 
  + \left( \frac{\asb}{2 \pi} \right)  {\cal A}_1^{b}(m^0,\eps) 
  + \left( \frac{\asb}{2 \pi}\right)^2  {\cal A}_2^{b}(m^0,\eps) +{\cal O}(\alpha_s^3)
  \right], \label{eq:bare_expansion}
\end{align}
where the dependence on kinematics is implicit.
The renormalised amplitude has the form
\begin{equation}
 {\cal A}^R\left(\as(\mu),y_t,m,\mu,\eps\right)
 = \left( \mu^2 \right)^{-\frac{3}{2}\eps}S_\eps^\frac{3}{2} \,
Z_q \, Z_Q \, {\cal A}^b
(S_\eps^{-1}\mu^{2\eps}Z_\alpha\as, S_\eps^{-\frac{1}{2}}\mu^{\eps}Z_my_t, Z_m m,\eps) \, ,
\end{equation}
where
\begin{equation}
    S_\eps\equiv\left(\frac{e^{\gamma_E}}{4\pi}\right)^{-\eps},
    \label{eq:seps}
\end{equation}
$Z_q$ ($Z_Q$) are the on-shell wave-function renormalisation
constants for light (heavy) quarks, $\mu$ is the renormalisation scale, and the bare quantities are replaced by the respective renormalised ones.
The bare mass of the heavy quark, $m^0$, is renormalised using $Z_m$ in the on-shell scheme and the bare Yukawa coupling $y_t^0$ is renormalised accordingly.
In particular, we further expand the coefficients of the expansion of the bare amplitude in \ref{eq:bare_expansion} to
\begin{align}
{\cal A}_n^{b}(Z_mm,\eps)&=m^{-1-n\eps}\bigg[{\cal A}_n(\eps)+\left(\aspi\right) Z_m^{(1)}{\cal A}_n^{(1)mct}(\eps)\\
                         &+\left(\aspi\right)^2\left(\frac{1}{2}(Z_m^{(1)})^2{\cal A}_n^{(2)mct}(\eps)+Z_m^{(2)}{\cal A}_n^{(1)mct}(\eps)\right)\bigg]+{\cal O}\left(\as^3\right)\;,
\end{align}
where ${\cal A}_n^{(k)mct}(\eps)=m^{1+n\eps}\left(m^k\partial_m^k{\cal A}_n^{b}(m,\eps)\right)$ and the overall factor of $m^{-1-n\eps}$ is extracted in order to have dimensionless amplitudes that depend on the dimensionless variables $x_{ij}$ introduced in \ref{eq:variable_def}.

The bare coupling constant $\asb$ is defined as
\begin{eqnarray}
\asb = S_\eps^{-1} \, \mu^{2\eps} \, Z^{(N_f)}_{\as} \as^{(N_f)}(\mu) \; ,
\end{eqnarray}
which corresponds to the $\MSbar$ scheme with $N_f = n_l+ n_h$ active flavours.
However, we do not consider the top quark as an active flavour contributing to the running of $\as$ and the parton distribution functions.
Therefore we use the decoupling relation
\begin{equation}
\as^{(N_f)} = \zeta_{\as} \as^{(n_l)} \; ,
\end{equation}
where $\zeta_{\as}$ is given in \ref{app:UV},
together with the explicit expressions for the renormalisation constants.
Our notation is such that $\as\equiv\as^{(N_f)}$.

To split two-loop amplitudes into smaller building blocks, it is in general convenient to project the amplitude onto scalar form factors multiplying the independent spinor and Lorentz structures, or onto helicity amplitudes.
The latter is however less convenient for amplitudes involving massive fermions.
For the $q\bar{q}$ channel considered here, we use the Born
amplitude itself as a projector, and calculate the spin- and
colour-summed interference term of the renormalised and rescaled
NNLO amplitude with the LO amplitude.
We decompose this quantity into colour and flavour factors as follows:
\begin{equation}
    \mathrm{Re}\braket{{\cal A}^R_0}{{\cal A}^R_2}\Big|_{N_f\text{-part}}=C_FN_CT_F \, {\cal C},
   \label{eq:interference2}
\end{equation}
\begin{align}\label{eq:amp2L}
    {\cal C} &= n_l^2T_F^2\,{\cal C}_{ll}+n_ln_hT_F^2\,{\cal C}_{lh}+n_h^2T_F^2\,{\cal C}_{hh}\nn\\
    &   +n_lC_FT_F\,{\cal C}_{lC_F}+n_lC_AT_F\,{\cal C}_{lC_A}+n_l\frac{d_{33}T_F}{C_FN_CT_F}\,{\cal C}_{ld_{33}}\nn\\
    &   +n_hC_FT_F\,{\cal C}_{hC_F}+n_hC_AT_F\,{\cal C}_{hC_A}+n_h\frac{d_{33}T_F}{C_FN_CT_F}\,{\cal C}_{hd_{33}} ,
\end{align}
where $N_C$ is the number of QCD colours, and the colour
group factors $C_F$, $C_A$, $T_F$, and $d_{33}$ are given in
Ref.~\cite{vanRitbergen:1998pn} for $SU(N)$ as well as $SO(N)$
and $Sp(N)$: we allow the colour group to be general here.

Similarly, the decomposition of the NLO and LO interference terms is
\begin{equation}
    \mathrm{Re}\braket{{\cal A}^R_0}{{\cal A}^R_1}=C_FN_CT_F \, {\cal B},
    \label{eq:interference1}
\end{equation}
\begin{equation}\label{eq:amp1L}
    {\cal B} = n_lT_F\,{\cal B}_{l}+n_hT_F\,{\cal B}_{h} +C_F\,{\cal B}_{C_F}+C_A\,{\cal B}_{C_A}+\frac{d_{33}}{C_FN_CT_F}\,{\cal B}_{d_{33}} \;,
\end{equation}
\begin{equation}
    \braket{{\cal A}^R_0}{{\cal A}^R_0} = C_FN_CT_F \,{\cal A}\;.
\label{eq:interference0}
\end{equation}
The explicit expressions for the renormalized components ${\cal A}$,
${\cal B}_i$ and ${\cal C}_i$ in terms of their bare counterparts
are given in \ref{eq:calA} to \ref{eq:calC3} in \ref{app:UV}.

Note that the colour factors of the $t\bar{t}H$ production amplitude
are in principle the same as for $t\bar{t}$ production.
The virtual amplitudes for top quark pair production at NNLO  were investigated in
Ref.~\cite{Baernreuther:2013caa}, see also Refs.~\cite{Bonciani:2008az,Bonciani:2009nb,Mandal:2022vju}
for analytic results in the quark channel. In Ref.~\cite{Baernreuther:2013caa}, the colour factor
decomposition is given after having formed the interference
with the Born amplitude, just as in \ref{eq:interference2}.
In contrast to~\cite[eq.~(2.8)]{Baernreuther:2013caa} however,
we do not assume the colour group to be $SU(N)$, and as a result,
instead of seven independent colour and flavour structures, we
identify nine.

\subsection{Infrared singularity structure of the virtual amplitude}
\label{sec:IRpoles}

The origin of the IR divergences present in 2-loop amplitudes with two massive coloured final state particles has been discussed in the literature and their form at 2-loops is known~\cite{Ferroglia:2009ii,Becher:2009kw,Baernreuther:2013caa}.

For the description of the IR divergences, we work in the colour space formalism. The renormalised amplitude is expressed as a vector in colour space $\ket{{\cal A}^{R}(\asl,y_t,m,\mu,\eps)}$ and the divergences are removed by using a multiplicative $\MSbar$ renormalisation factor ${\mathbf{Z}\equiv\mathbf{Z}(\{p\},\{m\},\mu,\eps)}$, which is a kinematic-dependent matrix in colour space.
For the colour decomposition for $q_{l}\bar{q}_{k}\to t_{i}\bar{t}_{j}H$ we adopt the following basis elements\footnote{This basis assumes the $SU(N)$ colour group rather than a more general colour group, %as the latter case would lead to a more complicated form of the vector $\ket{c_2}$. 
as the colour space in the latter case can be of higher dimension.  
However, since only the 1,1-component (upper left entry) of the colour matrix is needed for the $N_f$-part, our results are also valid for more general colour groups.}:
\begin{align}
    \ket{c_1}&=t^a_{ij}t^a_{kl}\;, & \ket{c_2}&=\delta_{ij}\delta_{kl}\;,
    \label{eq:colourbasis}
\end{align}
with full details given in \ref{app:Colour}.
We use the following expressions:
\begin{align}
\ket{{\cal A}^{fin}(\asl,y_t,m,\mu)}=\mathbf{Z}^{-1}\ket{{\cal A}^{R}(\asl,y_t,m,\mu,\eps)}\;.\label{eq:IR_removal}
\end{align}
The renormalisation constant $\mathbf{Z}$ fulfills the differential equation
\begin{align}
    \mu \frac{\d{}}{\d{}\mu} \mathbf{Z}(\{p\},\{m\},\mu,\eps)=-\mathbf{\Gamma}(\{p\},\{m\},\mu)\,\mathbf{Z}(\{p\},\{m\},\mu,\eps)\;,
\end{align}
which induces the following solution at two loops:
\begin{align}
   \mathbf{Z}
  &{}= 
    1 + \left(\aslpi\right)\left(\frac{\Gamma^{(1)\prime}}{4\eps^2}+\frac{\mathbf{\Gamma}^{(1)}}{2\eps}\right)
\nn\\
  &
   + \left(\aslpi\right)^2\left\{
    \frac{\left(\Gamma^{(1)\prime}\right)^2}{32\eps^4}
    + \frac{\Gamma^{(1)\prime}}{8\eps^3}\left(\mathbf{\Gamma}^{(1)}-\frac{3}{4}b_0\right)
    + \frac{\mathbf{\Gamma}^{(1)}}{8\eps^2}\left(\mathbf{\Gamma}^{(1)}-b_0\right)
    + \frac{\Gamma^{(2)\prime}}{16\eps^2}
    + \frac{\mathbf{\Gamma}^{(2)}}{4\eps}
   \right\}\;.
\end{align}
The coefficients of the anomalous dimension matrix are defined by the expansion
\begin{equation}
    \mathbf{\Gamma}=\left(\aslpi\right)\mathbf{\Gamma}^{(1)}+\left(\aslpi\right)^2\mathbf{\Gamma}^{(2)},
\end{equation}
with $\Gamma^{\prime}=\mu \frac{\partial{}}{\partial{}\mu} \Gamma$ and $b_0=\left[\beta_0\right]_{N_f\to n_l}$.
The general form of the anomalous dimension matrix up to two loops is given in Ref.~\cite{Ferroglia:2009ii}.
Here we present the explicit form of the expressions for the process we study.
With the colour basis of \ref{eq:colourbasis} the anomalous
dimension matrix has the specific form
\begin{align}
 \mathbf{\Gamma}=
 &\left(C_F\left[ \gamma_{cusp}\!\left(\asl\right)\log \left( \alpha_1 \right)+\gamma_{cusp}\!\left(\beta_{34},\asl\right)\right]+2\gamma_q\!\left(\asl\right)+2\gamma_Q\!\left(\asl\right)\right)
 \begin{pmatrix}
1 & 0\\
0 & 1 
 \end{pmatrix}
 \nn\\
&+\left(\frac{C_A}{4}\left[\gamma_{cusp}\!\left(\asl\right)\log \left(\alpha_2\right) -2\gamma_{cusp}\!\left(\asl\right)\log \left(\alpha_1\right) -2\gamma_{cusp}\!\left(\beta_{34},\asl\right)\right]
\right.\nn\\
&\quad\left.+\frac{d_{33}}{T_FC_FN_C}\gamma_{cusp}\!\left(\asl\right)\log \left( \alpha_3 \right) \right)
 \begin{pmatrix}
1 & 0\\
0 & 0 
 \end{pmatrix}
 \nn\\
 &+\gamma_{cusp}\!\left(\asl\right)\log \left (\alpha_3 \right) \left(
  \begin{pmatrix}
0 & 1\\
\frac{T_FC_F}{N_C} & 0 
 \end{pmatrix}
 +C_A\frac{1}{2}\aslpi g\!\left(\beta_{34}\right)
 \begin{pmatrix}
0 & -1\\
\frac{T_FC_F}{N_C} & 0 
 \end{pmatrix}\right)
 \nn\\
  &+{\cal O}\left((\asl)^3\right)\;,
    \label{eq:Gamma}
\end{align}
with
\begin{align}
&\alpha_1 = \frac{-sp_{12}}{\mu^2},&
&\alpha_2 = \frac{sp_{13}\,sp_{14}\,sp_{23}\,sp_{24}}{m^4\mu^4},&
&\alpha_3 = \frac{sp_{13}\,sp_{24}}{sp_{23}\,sp_{14}},&
&\beta_{34} = \mathrm{acosh}\frac{-sp_{34}}{2m^2},&
\end{align}
and $sp_{ij} \equiv 2 p_i\cdot p_j + i 0^+$. The anomalous dimensions $\gamma_i$ are given in \ref{app:IR}.
Since we are only interested in the interference with the LO amplitude, we only need the component $\Gamma_{11}$ of the anomalous dimension matrix $\mathbf{\Gamma}$ for the IR pole structure of those amplitude parts proportional to the quark flavours at NNLO. Expanding in $\asl$ and in terms of light flavour and colour factors  leads to
\begin{align}
\Gamma_{11}&{}=\frac{\asl}{2\pi}\left(C_F\Gamma_{11,\;C_F}^{(1)}+C_A\Gamma_{11,\;C_A}^{(1)}+\frac{d_{33}}{C_FN_CT_F}\Gamma_{11,\;d_{33}}^{(1)}\right)
\nn\\
&+\left(\frac{\asl}{2\pi}\right)^2\left(n_lC_FT_F\Gamma_{11,\;n_lC_F}^{(2)}+n_lC_AT_F\Gamma_{11,\;n_lC_A}^{(2)}+n_l\frac{d_{33}T_F}{C_FN_CT_F}\Gamma_{11,\;n_ld_{33}}^{(2)}+\dots\right)
\nn\\
&+{\cal O}\left((\asl)^3\right) \;,
\nn\\
\Gamma_{11}^\prime&{}=\aslpi \left(-2C_F\gamma_{cusp}^{(1)}\right)+\left(\aslpi\right)^2 \left(-2C_FC_A\gamma_{cusp,C_A}^{(2)}-2n_lC_FT_F\gamma_{cusp,n_l}^{(2)}\right)\;,
\end{align}
where the parts irrelevant for the $N_f$-part of the amplitude are contained in the dots.
The relevant $Z_{11}$ components then have the form
\begin{align}
  \label{eq:Z11}
Z_{11}&{}=1 + \left(\aslpi\right)\left[C_F\left(\frac{-2\gamma_{cusp}^{(1)}}{4\eps^2}+\frac{\Gamma_{11,\;C_F}^{(1)}}{2\eps}\right)+C_A\frac{\Gamma_{11,\;C_A}^{(1)}}{2\eps}+\frac{d_{33}}{C_FN_CT_F}\frac{\Gamma_{11,\;d_{33}}^{(1)}}{2\eps}\right]
\nn\\
&
+ \left(\aslpi\right)^2\left[n_lC_FT_F\left(\frac{-2\gamma_{cusp}^{(1)}}{8\eps^3}+\frac{\Gamma_{11,\;C_F}^{(1)}}{6\eps^2}+\frac{-2\gamma_{cusp,n_l}^{(2)}}{16\eps^2}+\frac{\Gamma_{11,\;n_lC_F}^{(2)}}{4\eps}\right)
\right.\nn\\
&\left.\quad
+n_lC_AT_F\left(\frac{\Gamma_{11,\;C_A}^{(1)}}{6\eps^2}+\frac{\Gamma_{11,\;n_lC_A}^{(2)}}{4\eps}\right)
\right.\nn\\
&\left.\quad+n_l\frac{d_{33}T_F}{C_FN_CT_F}\left(\frac{\Gamma_{11,\;d_{33}}^{(1)}}{6\eps^2}+\frac{\Gamma_{11,\;n_ld_{33}}^{(2)}}{4\eps}\right)+\dots\right] \;,
\end{align}
which can now be used in order to construct the IR structure of the interference terms, as will be shown in the following:
\begin{align}
    {\cal B}_{C_F}^\text{IR}={}&\left(\frac{-2\gamma_{cusp}^{(1)}}{4\eps^2}+\frac{\Gamma_{11,\;C_F}^{(1)}}{2\eps}\right) {\cal A}\;, \nn\\
    {\cal B}_{C_A}^\text{IR}={}&\frac{\Gamma_{11,\;C_A}^{(1)}}{2\eps}\,{\cal A}\;,\quad
    {\cal B}_{d_{33}}^\text{IR}=\frac{\Gamma_{11,\;d_{33}}^{(1)}}{2\eps}\,{\cal A}\;,
\end{align}
\begin{align}
{\cal C}_{lC_F}^\text{IR}=&\left(\frac{-2\gamma_{cusp}^{(1)}}{8\eps^3}+\frac{\Gamma_{11,\;C_F}^{(1)}}{6\eps^2}+\frac{-2\gamma_{cusp,n_l}^{(2)}}{16\eps^2}+\frac{\Gamma_{11,\;n_lC_F}^{(2)}}{4\eps}\right) {\cal A}
+\left(\frac{-2\gamma_{cusp}^{(1)}}{4\eps^2}+\frac{\Gamma_{11,\;C_F}^{(1)}}{2\eps}\right) {\cal B}_l \;,
\nn\\
{\cal C}_{lC_A}^\text{IR}=&\left(\frac{\Gamma_{11,\;C_A}^{(1)}}{6\eps^2}+\frac{\Gamma_{11,\;n_lC_A}^{(2)}}{4\eps}\right) {\cal A}
+\frac{\Gamma_{11,\;C_A}^{(1)}}{2\eps}\,{\cal B}_l \;,
\nn\\
{\cal C}_{ld_{33}}^\text{IR}=&\left(\frac{\Gamma_{11,\;d_{33}}^{(1)}}{6\eps^2}+\frac{\Gamma_{11,\;n_ld_{33}}^{(2)}}{4\eps}\right) {\cal A}
+\frac{\Gamma_{11,\;d_{33}}^{(1)}}{2\eps}\,{\cal B}_l \;,
\end{align}
\begin{align}
{\cal C}_{hC_F}^\text{IR}=&\left(\frac{-2\gamma_{cusp}^{(1)}}{4\eps^2}+\frac{\Gamma_{11,\;C_F}^{(1)}}{2\eps}\right) {\cal B}_h \;,
\nn\\
{\cal C}_{hC_A}^\text{IR}=&\frac{\Gamma_{11,\;C_A}^{(1)}}{2\eps}\,{\cal B}_h\;,\quad
{\cal C}_{hd_{33}}^\text{IR}=\frac{\Gamma_{11,\;d_{33}}^{(1)}}{2\eps}\,{\cal B}_h\;.
\end{align}
The parts not shown here have no IR poles.

\subsection{Workflow of the calculation}
%%%%%%%%%%%%%%%%%%%%%

The leading order (LO) amplitude $\mathcal{A}_0^b$ can be
represented by two Feynman diagrams:
\begin{equation}
    \raisebox{0.5ex}{\scalebox{0.75}{\input{figs/qqtth-1.tikz}}} \qquad\text{and}\qquad \raisebox{0.5ex}{\scalebox{0.75}{\input{figs/qqtth-2.tikz}}}.
\end{equation}
The LO amplitude has no $N_f$-part itself, but it contributes
to the renormalisation of the NNLO $N_f$-part, because the
$\alpha_s$ beta-function contains $N_f$.
The LO amplitude in the quark channel has both $\eps^0$ and $\eps^1$ parts (but no higher parts).
We derive the corresponding expression using \textsc{Alibrary}~\cite{alibrary},
which is a Mathematica library interfacing with
\textsc{Qgraf}~\cite{Nogueira:1991ex}, \textsc{Feynson}~\cite[Chapter~4]{Magerya:2022esf},
\textsc{Form}~\cite{Ruijl:2017dtg}, and \textsc{Color.h}~\cite{vanRitbergen:1998pn}
to generate amplitudes, sum over tensor structures,
construct integral families, and export the results
to integration-by-parts~(IBP) relation solvers and/or
\pysecdec{}~\cite{Heinrich:2023til,Heinrich:2021dbf,Borowka:2018goh,Borowka:2017idc}.

\begin{figure}
    \centering
    \includegraphics[width=\textwidth]{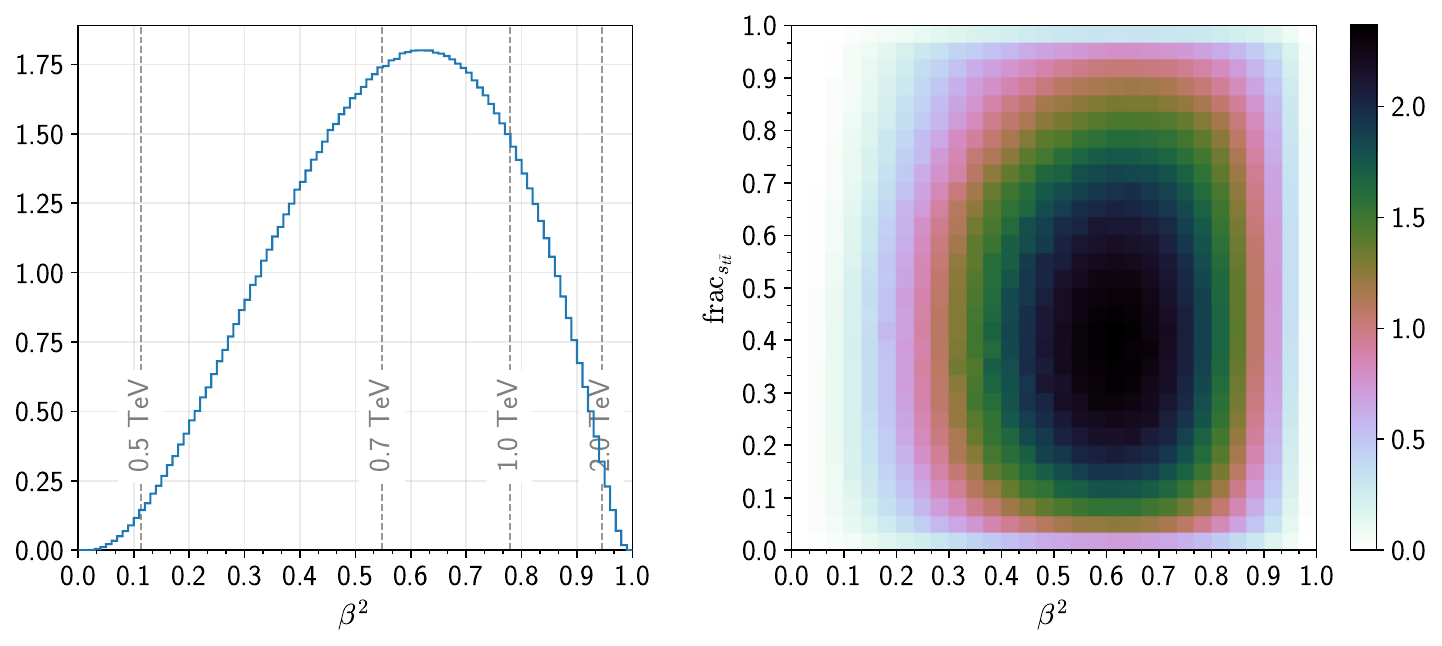}
    \caption{%
        Event probability distribution in $\beta^2$ (left),
        and $\beta^2$ and $\fstt$ (right), according to the LO
        $q\bar{q}\to t\bar{t}H$ amplitude.
        For this plot we take the energy of incoming quarks to be
        distributed according to the ABMP16 parton distribution
        functions~\cite{Alekhin:2017kpj} (which we evaluate
        via LHAPDF~\cite{Buckley:2014ana}), with the collision
        energy set to 13.6~TeV.
        We have also applied cuts on the top quark momenta (as
        we calculate with on-shell top-quarks) in line with
        those reported in \cite{Aaboud:2018urx,CMS:2018fdh}:
        we enforce a minimal transverse momentum of 25~GeV, a
        maximal rapidity of~4.5, and a separation $\Delta R$ in
        rapidity and azimuthal angle between the top quarks of
        $\Delta R>0.4$.
        These cuts remove about 3\% of the events, and mostly
        affect the low-$\beta$ region.
    }
    \label{fig:event-distribution}
\end{figure}

We can use the LO result to
estimate the distribution of the events over the phase space at
the LHC, as done in \ref{fig:event-distribution}.
These plots tell us that most of the events are expected to come
from the region of moderately high $\beta^2$ and medium $\fstt$.
In particular, the region of $\beta^2\in[0.10, 0.95]$ (that is,
$\sqrt{\hat{s}}\in{}$[500~GeV, 2.1~TeV]) is expected to contain 99\%
of all events.

\subsubsection{Amplitude generation}

\input{diagrams1l.tex}

\input{diagrams2l.tex}

To generate the one-loop and two-loop amplitudes ($\mathcal{A}^b_1$ and $\mathcal{A}^b_2$
respectively) we use the following procedure:
first we generate the corresponding Feynman
diagrams (using \textsc{Qgraf}), then we insert Feynman rules,
apply the projectors, and sum over the spinor and colour tensors
(using \textsc{Form} and \textsc{Color.h}); all of this is done
through \textsc{Alibrary}.
This way, for each diagram, we obtain a corresponding sum of many
scalar integrals.

In total we find 31 non-zero one-loop diagrams and 249 two-loop diagrams.
Examples for one-loop diagrams with different colour factors are
depicted in \ref{fig:example_diagrams1l}; examples for two-loop diagrams
can be found in \ref{fig:example_diagrams2l}.\footnote{These diagrams
were prepared using \textsc{FeynGame}~\cite{Harlander:2020cyh,Harlander:2024qbn}.}

\subsubsection{IBP reduction}

The next step is to reduce the calculation of the approximately 20000 scalar integrals that appear in
the amplitudes to a much smaller number of master integrals using IBP relations~\cite{Chetyrkin:1981qh}.
To this end we first calculate the symmetries between the diagrams (using \textsc{Feynson}), and sort them into integral families.
For the $q\bar{q}$ channel we use 4 one-loop and 43 two-loop families.
Out of the two-loop families, 28 are unique (shown in~\ref{app:familyplots}); the remaining 15 are obtained by crossing the external momenta.

The usual next step would be to write down a system of IBP
equations and solve it symbolically using e.g.\ the Laporta
algorithm~\cite{Laporta:2000dsw}.
We employed \textsc{Kira}~\cite{Maierhofer:2017gsa, Klappert:2020nbg} together
with \textsc{Firefly}~\cite{Klappert:2019emp, Klappert:2020aqs} for this purpose and observed that
while the reductions for the one-loop families may be obtained in a rather straightforward manner,
for most of the two-loop families the computation is quite challenging.
Fully analytic reduction of the two-loop amplitude is rendered largely unfeasible given the large number of variables (five ratios given in~\ref{eq:variable_def}, the mass ratio $m_H^2/m_t^2$, and the dimensional regulator~$\eps$) along with the presence of internal masses.
Instead we opt for a numerical approach and solve the IBP systems for individual phase-space points by substituting kinematic scales with rational numbers.
Note that we employ the same numerical approach for one-loop amplitudes as well so as to have a uniform implementation for the whole calculation.

To set up the IBP reduction, we first select a basis of master integrals for each of the amplitudes.
We require a total of 33 master integrals for the one-loop amplitudes and 831 master integrals for the two-loop amplitudes.
The choice of master integrals significantly impacts both amplitude reduction as well as numerical evaluation of the master integrals.
Ideally we prefer a basis that is
1)~quasi-finite~\cite{vonManteuffel:2014qoa},
2)~$d$-factorising~\cite{Smirnov:2020quc,Usovitsch:2020jrk},
3)~fast to evaluate with \pysecdec, and
4)~leads to simple polynomials in the denominators of the IBP reduction tables.
Finding a basis satisfying the first two requirements is rather straightforward by considering integrals in higher dimensions ($d=6-2\eps$ or $d=8-2\eps$)
and with higher propagator powers or dots, with up to 5 dots in some cases.
We then apply heuristic arguments to choose integrals that also satisfy the last two requirements in the following way.
For each sector, we perform a reduction neglecting sub-sector integrals,
and we analyze the denominator factors of the resulting IBP tables 
for different choices of master integrals. 
Selecting the master integral basis with the smallest denominator factors,
we observe a significant improvement in the run-time for the full reduction.
With this initial basis choice, we then evaluate the amplitude as discussed below
and we identify the integrals with a significant contribution to the evaluation time. 
The basis of the corresponding sectors is then further refined by 
repeating the above procedure, restricting the set of candidate masters
to integrals showing a relatively fast convergence with \pysecdec.

After we select the basis, we use \textsc{Kira} to generate the IBP systems for each integral family.
We generate dimensional recurrence relations using \textsc{Alibrary} to be able to reduce the amplitudes to master integrals in shifted dimensions.
The combined system of equations is then fed to \textsc{Ratracer}~\cite{Magerya:2022hvj} which prepares and optimizes an \textit{execution trace}
of the solution.
Then we use \textsc{Ratracer} to perform a series expansion on this trace in~$\eps$; this results in direct output of the $\eps$-expansion of the IBP coefficients.
Performing an expansion in $\eps$ effectively removes it from the computation which, combined with substituting rational numbers for the kinematic scales, 
results in a purely numerical system.
This system is then solved by \textsc{Ratracer} through replaying
the trace in a parallelized manner and using finite field methods.
Note that finite field methods used for function reconstruction as
a way of solving IBP equations is by now an established practice,
pioneered in Refs.~\cite{vonManteuffel:2014ixa,Peraro:2016wsq};
our usage however does not require function reconstruction,
only rational number reconstruction and the Chinese remainder
theorem.
Our setup allows us to compute reductions in around two CPU
minutes for the two-loop amplitude, and under a second for the
one-loop amplitude on a desktop CPU for most points.
Overall this reduction method is fast enough, in the sense that we are more
constrained by the evaluation of the master integrals.

\subsubsection{Evaluation of the master integrals}

The families of integrals required in this calculation are
complicated enough that analytic evaluation is not currently
achievable for many of the master integrals, such as topologies b81 or
b82 shown in \ref{app:familyplots}.
Instead, we turn to evaluating the master integrals numerically, using the approach of
sector decomposition as implemented in \pysecdec{}.
Specifically, we rely on \pysecdec{}'s ability to integrate weighted
sums of integrals (introduced in version~1.5, see~\cite{Heinrich:2021dbf}),
and define one sum for each of the bare amplitude's symbolic structures
as given in \ref{eq:interference2} and \ref{eq:interference1}.
We require the two-loop amplitudes to be expanded up to $\eps^0$,
and one-loop up to $\eps^1$.

By default \pysecdec{} expects the coefficient of the sums to be
given as algebraic expressions in terms of kinematic variables,
but because we do not compute these in general (as we do not
perform a fully symbolic IBP reduction), we instead use the results
of a test IBP reduction at some fixed kinematic point for the
coefficients to compile the \pysecdec{} integration library.
This ensures that the pole structure of the amplitude is known at
the compilation stage, and so the needed depths of the $\eps$-expansion
of the masters can be correctly determined.
After the integration libraries are compiled, to evaluate
the amplitudes at a given kinematic point, we substitute the
coefficients with the results of the IBP reduction.

The sector decomposition of the 831 two-loop master integrals
results in a total of around 18000 sectors, and around 28000
sector/expansion-order pairs.
To make the evaluation of such a lage number of expressions efficient we rely on
the performance improvements in \pysecdec{}~1.6 (see~\cite{Heinrich:2023til})
coming from the new \textsc{Disteval} evaluator (which was
partially developed as a response to the challenges of this
calculation).
We perform all the evaluations of the two-loop amplitudes on
NVidia~A100 GPUs.

The one-loop amplitudes on the other hand are much simpler (180 sectors
in total) and quicker to evaluate; for them we only use CPUs.

Our target is to obtain the renormalised two-loop amplitudes with
a precision better than~1\%.
In the bulk of the phase space this is easily achievable, and
the two-loop integration time per point is around 5~minutes; for the
one-loop amplitude it is
around 10 seconds using 4~CPU threads.
This however changes in the high-$\beta$ region and in the regions
near the boundaries of $\fstt$ and $\theta_t$: there we observe large
numerical cancellations, both within and between the integrals,
that require the evaluation of the master integrals to higher
precisions to meet the amplitude precision goals.

These cancellations cause three separate problems:
\begin{enumerate}
\item
They drive the evaluation times upward, and in principle we
expect this growth to be unbounded as $\beta$ tends to~1
(i.e.\ $\hat{s}\to\infty$).

This problem could be mitigated by an 
 asymptotic expansion in large $\hat{s}$ for the high-$\beta$ region.
However, we will not follow this strategy here, in favour of 
using a single method for all regions.

\item
They require increasingly large Quasi-Monte-Carlo lattice sizes,
to an extent where we run into the limitations of precomputed lattices
available in \pysecdec{}: the largest such lattice has the size
of around $7\!\cdot\!10^{10}$, but some of the integrals need
up to $10^{14}$ evaluations in the high-$\beta$ region.

To solve this issue we employ  the new ``median QMC
rules''~\cite{GE22} lattice construction method implemented in
\pysecdec{}~1.6, that enables on-the-fly construction of lattices
of unbounded size.

\item
At very high $\beta^2$ (e.g.\ 0.99) the cancellations between
some of the integrals become as large as 20 decimal digits, which
means that even evaluating the integrals to the full precision of
the double-precision floating point numbers (which is 16 digits)
would be insufficient to get any precision for the amplitudes.

Here we find that the integrals that cancel between each other
and need high precision are mostly simpler integrals with up to
five denominators, most significantly the ones of ``sunset'' and 
``ice-cone''\,\cite{Duhr:2022dxb} type, in various mass and external momenta configurations.
Such integrals converge relatively quickly, and obtaining them
with more than 20 digits of precision using sector decomposition
would be well within our time budget if it was not for the
double-precision floating point limitation.

\end{enumerate}

As a solution we have upgraded \pysecdec{} with the ability
to dynamically switch from double floating precision to
``double-double'' for integrals that need it, allowing for the
maximum of 32 decimal digits of precision.
Our implementation of the double-double arithmetic is based on
the methods described in~\cite{Bailey2003,Shewchuk97}.
We choose this approach instead of the more commonly used
quadruple precision floating point numbers (\texttt{float128})
because NVidia GPU compilers do not come with the support for
either of them, and our benchmarks show that double-double performs
around 2.5~times faster than \texttt{float128} on a CPU,
while being simple enough to be implemented on a GPU.
Still, the performance penalty of double-double integration is
as high as a factor of 20 on the GPU compared to double.

To cross-check our double-double precision implementation
we have also evaluated the sunset and ice-cone type integrals 
using the series solution of differential equations as
implemented in the DiffExp package~\cite{Moriello:2019yhu,
Hidding:2020ytt} with boundary conditions obtained using the 
Auxiliary Mass Flow (AMFlow) method~\cite{Liu:2022chg}.
We find agreement between our results up to
the error reported by our double-double implementation.

%\end{enumerate}

Once the integrals are evaluated, the last step is to combine the
values of the bare Born, one-loop and two-loop results to values for the renormalised virtual
two-loop amplitude as described in \ref{sec:uv-renormalisation},
propagating the numerical uncertainties.

\subsection{Checks}

To double-check our calculation we have independently computed the LO
and NLO amplitudes via \gosam{}~\cite{Cullen:2011ac,Cullen:2014yla}
at a number of points, verifying agreement within the reported
accuracy.
Note that the comparison of the NLO amplitudes requires extra care,
because \gosam{} produces the results in the 't~Hooft-Veltman
scheme~\cite{tHooft:1972tcz}, and these need to be converted to
get agreement with conventional dimensional regularisation that
we use.
Regularisation scheme independence of the NLO virtual contribution is only obtained 
after full IR subtraction~\cite{Catani:1996pk}.
In the particular case of interest, 
the scheme difference of the subtraction term can be traced to the $\mathcal{O}(\eps^1)$ part of the Born contribution which vanishes for 
the 't~Hooft-Veltman scheme. 
Hence, we can convert the \gosam{} result to conventional dimensional regularisation by
\begin{equation}
    \mathrm{Re}\braket{{\cal A}^R_0}{{\cal A}^R_1}=\mathrm{Re}\braket{{\cal A}^R_0}{{\cal A}^R_1}^\text{\gosam{}}\left(1-\frac{\pi^2}{12}\eps^2\right)
    +Z_{11}^{(1)} \mathcal{A}\Big|_{\eps^1\text{-part}} +\mathcal{O}(\eps)\;.
\end{equation}
The factor $1-\frac{\pi^2}{12}\eps^2$ is necessary because the convention for $\MSbar$ in \gosam{}
uses $\frac{(4\pi)^\eps}{\Gamma(1-\eps)}$ instead of $S_\eps$ of \ref{eq:seps} as a prefactor.

We have also double-checked the IR poles of our amplitudes
against Ref.~\cite{Chen:2022nxt}, where the pole parts of the
renormalised interference terms are given at four phase-space
points.
To get agreement with this paper we need to set the renormalisation
scale to $m_t$, and use 6 fermion flavours in the running of
$\alpha_s$.

While some of the symmetries listed in \ref{sec:symmetries} are
trivially observed when deriving the variables of \ref{eq:variable_def}
from the parameters of \ref{eq:parameters}, we have verified the
symmetry of simultaneous exchange of $q\leftrightarrow\bar{q}$
and $t\leftrightarrow\bar{t}$.

Finally and most importantly, for each evaluated point we
compute the predicted IR pole coefficients of the amplitude as described
in \ref{sec:IRpoles}, and compare them to the ones obtained by
numerical integration.
We find agreement within the reported integration accuracy,
which provides us with a check on both the correctness of the
renormalisation procedure, and on the correctness of the
reported numerical integration precision of the two-loop results, since the
IR poles only depend on the one-loop amplitudes, which are integrated
separately (and to a higher precision).

%% file: diagrams1l.tex
\begin{figure}
\centering

\begin{subfigure}[c][][c]{0.32\textwidth}
\begin{tabular}{c}\def\svgwidth{1.\textwidth}\input{./diagrams/1loop/Diagram1.tex}\end{tabular}
% \caption{Projector 1: $n_h T_F$}
\caption{$n_h T_F$}
\end{subfigure}
\begin{subfigure}[c][][c]{0.32\textwidth}
\begin{tabular}{c}\def\svgwidth{1.\textwidth}\input{./diagrams/1loop/Diagram2.tex}\end{tabular}
% \caption{Projector 1: $C_A$}
\caption{$C_A$}
\end{subfigure}

\begin{subfigure}[c][][c]{0.32\textwidth}
\begin{tabular}{c}\def\svgwidth{1.\textwidth}\input{./diagrams/1loop/Diagram3.tex}\end{tabular}
% \caption{Projector 1: $C_A - 2 C_F$}
\caption{$C_A - 2 C_F$}
\end{subfigure}
\begin{subfigure}[c][][c]{0.32\textwidth}
\begin{tabular}{c}\def\svgwidth{1.\textwidth}\input{./diagrams/1loop/Diagram4.tex}\end{tabular}
% \caption{Projector 1: $C_F$}
\caption{$C_F$}
\end{subfigure}
\begin{subfigure}[c][][c]{0.32\textwidth}
\begin{tabular}{c}\def\svgwidth{1.\textwidth}\input{./diagrams/1loop/Diagram5.tex}\end{tabular}
% \caption{Projector 1: $n_l T_F$}
\caption{$n_l T_F$}
\end{subfigure}
\begin{subfigure}[c][][c]{0.32\textwidth}
\begin{tabular}{c}\def\svgwidth{1.\textwidth}\input{./diagrams/1loop/Diagram6.tex}\end{tabular}
% \caption{Projector 1: $\frac{C_A^2 C_F T_F - 4 d_{33}}{C_A C_F T_F}$}
\caption{$\frac{C_A C_F N_c T_F - 4 d_{33}}{C_F N_c T_F}$}
\end{subfigure}
\begin{subfigure}[c][][c]{0.32\textwidth}
\begin{tabular}{c}\def\svgwidth{1.\textwidth}\input{./diagrams/1loop/Diagram7.tex}\end{tabular}
% \caption{Projector 1: $\frac{C_A^2 C_F T_F + 4 d_{33}}{C_A C_F T_F}$}
\caption{$\frac{C_A C_F N_c T_F + 4 d_{33}}{C_F N_c T_F}$}
\end{subfigure}
%
% \begin{subfigure}[c][][c]{0.32\textwidth}
% \begin{tabular}{c}\def\svgwidth{1.\textwidth}\input{./diagrams/1loop/Diagram8.tex}\end{tabular}
% % \caption{Projector 2: $\frac{C_F T_F}{C_A}$}
% \caption{Projector 2: $\frac{C_F T_F}{N_c}$}
% \end{subfigure}
%

\caption{%
    Example diagrams for $q\bar{q}\to t\bar{t}H$ at one-loop level.
    Massive quarks are depicted using solid (blue) bold lines, while massless quarks are represented by lighter (grey/red) solid lines.
    The colour factors correspond to applying the first colour
    projector from \ref{eq:colourbasis}.
    \label{fig:example_diagrams1l}}
\end{figure}

%% file: diagrams2l.tex
\begin{figure}
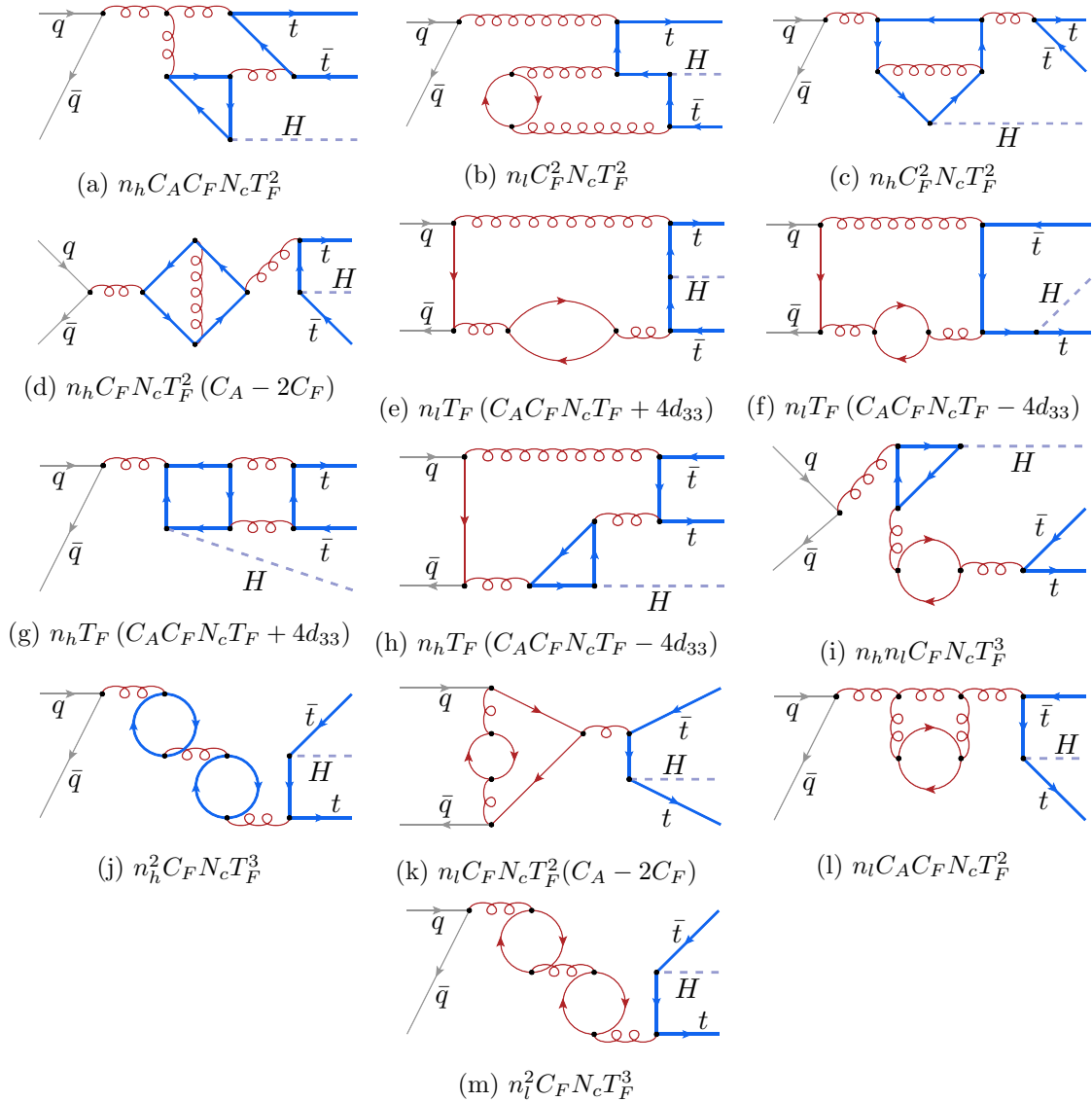

\centering

\begin{subfigure}[c][][c]{0.32\textwidth}
\begin{tabular}{c}\def\svgwidth{1.\textwidth}\input{./diagrams/2loop/Diagram1.tex}\end{tabular}
% \caption{$n_h C_A^2 C_F   T_F^2$}
\caption{$n_h C_A   C_F   N_c   T_F^2$}
\end{subfigure}
\begin{subfigure}[c][][c]{0.32\textwidth}
\begin{tabular}{c}\def\svgwidth{1.\textwidth}\input{./diagrams/2loop/Diagram2.tex}\end{tabular}
% \caption{$n_l C_A   C_F^2 T_F^2$}
\caption{$n_l       C_F^2 N_c   T_F^2$}
\end{subfigure}
\begin{subfigure}[c][][c]{0.32\textwidth}
\begin{tabular}{c}\def\svgwidth{1.\textwidth}\input{./diagrams/2loop/Diagram3.tex}\end{tabular}
% \caption{$n_h C_A   C_F^2 T_F^2$}
\caption{$n_h       C_F^2 N_c   T_F^2$}
\end{subfigure}
\begin{subfigure}[c][][c]{0.32\textwidth}
\begin{tabular}{c}\def\svgwidth{1.\textwidth}\input{./diagrams/2loop/Diagram4.tex}\end{tabular}
% \caption{$n_h C_A   C_F   T_F^2 \left( C_A - 2 C_F \right)$}
\caption{$n_h       C_F   N_c   T_F^2 \left( C_A - 2 C_F \right)$}
\end{subfigure}
\begin{subfigure}[c][][c]{0.32\textwidth}
\begin{tabular}{c}\def\svgwidth{1.\textwidth}\input{./diagrams/2loop/Diagram5.tex}\end{tabular}
% \caption{$n_l T_F \left( T_F C_A^2 C_F + 4 d_{33} \right)$}
\caption{$n_l T_F \left( C_A C_F N_c T_F + 4 d_{33} \right)$}
\end{subfigure}
\begin{subfigure}[c][][c]{0.32\textwidth}
\begin{tabular}{c}\def\svgwidth{1.\textwidth}\input{./diagrams/2loop/Diagram6.tex}\end{tabular}
% \caption{$n_l T_F \left( T_F C_A^2 C_F - 4 d_{33} \right)$}
\caption{$n_l T_F \left( C_A C_F N_c T_F - 4 d_{33} \right)$}
\end{subfigure}
\begin{subfigure}[c][][c]{0.32\textwidth}
\begin{tabular}{c}\def\svgwidth{1.\textwidth}\input{./diagrams/2loop/Diagram7.tex}\end{tabular}
% \caption{$n_h T_F \left( T_F C_A^2 C_F + 4 d_{33} \right)$}
\caption{$n_h T_F \left( C_A C_F N_c T_F + 4 d_{33} \right)$}
\end{subfigure}
\begin{subfigure}[c][][c]{0.32\textwidth}
\begin{tabular}{c}\def\svgwidth{1.\textwidth}\input{./diagrams/2loop/Diagram8.tex}\end{tabular}
% \caption{$n_h T_F \left( T_F C_A^2 C_F - 4 d_{33} \right)$}
\caption{$n_h T_F \left( C_A C_F N_c T_F - 4 d_{33} \right)$}
\end{subfigure}
\begin{subfigure}[c][][c]{0.32\textwidth}
\begin{tabular}{c}\def\svgwidth{1.\textwidth}\input{./diagrams/2loop/Diagram9.tex}\end{tabular}
% \caption{$n_h n_l C_A C_F T_F^3$}
\caption{$n_h n_l C_F N_c T_F^3$}
\end{subfigure}
\begin{subfigure}[c][][c]{0.32\textwidth}
\begin{tabular}{c}\def\svgwidth{1.\textwidth}\input{./diagrams/2loop/Diagram10.tex}\end{tabular}
% \caption{$n_h^2   C_A C_F T_F^3$}
\caption{$n_h^2 C_F N_c T_F^3$}
\end{subfigure}
\begin{subfigure}[c][][c]{0.32\textwidth}
\begin{tabular}{c}\def\svgwidth{1.\textwidth}\input{./diagrams/2loop/Diagram11.tex}\end{tabular}
% \caption{$n_l C_A   C_F   T_F^2 \left( C_A - 2 C_F \right)$}
\caption{$n_l C_F N_c T_F^2 (C_A - 2 C_F)$}
\end{subfigure}
\begin{subfigure}[c][][c]{0.32\textwidth}
\begin{tabular}{c}\def\svgwidth{1.\textwidth}\input{./diagrams/2loop/Diagram12.tex}\end{tabular}
% \caption{$n_l C_A^2 C_F T_F^2$}
\caption{$n_l C_A C_F N_c T_F^2$}
\end{subfigure}
\begin{subfigure}[c][][c]{0.32\textwidth}
\begin{tabular}{c}\def\svgwidth{1.\textwidth}\input{./diagrams/2loop/Diagram13.tex}\end{tabular}
% \caption{$n_l^2 C_A C_F T_F^3$}
\caption{$n_l^2 C_F N_c T_F^3$}
\end{subfigure}

\caption{%
    Example diagrams for $q\bar{q}\to t\bar{t}H$ at two-loop level proportional to $n_l$ or $n_h$.
    Massive quarks are depicted using solid (blue) bold lines, while massless quarks are represented by lighter (grey/red) solid lines.
    The colour factors correspond to applying the first colour
    projector from \ref{eq:colourbasis}.
    \label{fig:example_diagrams2l}}
\end{figure}

%% file: results.tex
In this section we visualise the two-loop amplitude as a function of slices of phase-space variables.
To this end we choose the following kinematic point to centre our slices on:
\begin{equation}
    \beta^2 = 0.8, \;
    \fstt = 0.7, \;
    \cos \theta_H = 0.8, \;
    \cos \theta_t = 0.9, \;
    \cos \varphi_t = 0.7;
    \label{eq:centre}
\end{equation}
we also set $m_H^2/m_t^2 = 12/23$, $\mu^2=s_{12}/4$, and $\mt^2=1$.
The values of the amplitude at this kinematic point is given in
\ref{app:pspoint} (along with two other points), where we list both
the bare and the renormalised values of each component of the
LO, one- and two-loop amplitude (as defined in \ref{eq:amp2L},
\ref{eq:amp1L}, and \ref{eq:interference0}).
For brevity however we prefer not to plot individual
components, but rather the combined ${\cal C}$ and ${\cal B}$ values
as defined in \ref{eq:interference2} and \ref{eq:interference1}.
For this we set $n_l=5$, $n_h=1$, and the colour group to
$SU(3)$, i.e.\ $C_F=4/3$, $C_A=3$, and $d_{33}=5/6$.

\begin{figure}
  \centering
  \includegraphics[width=0.49\textwidth]{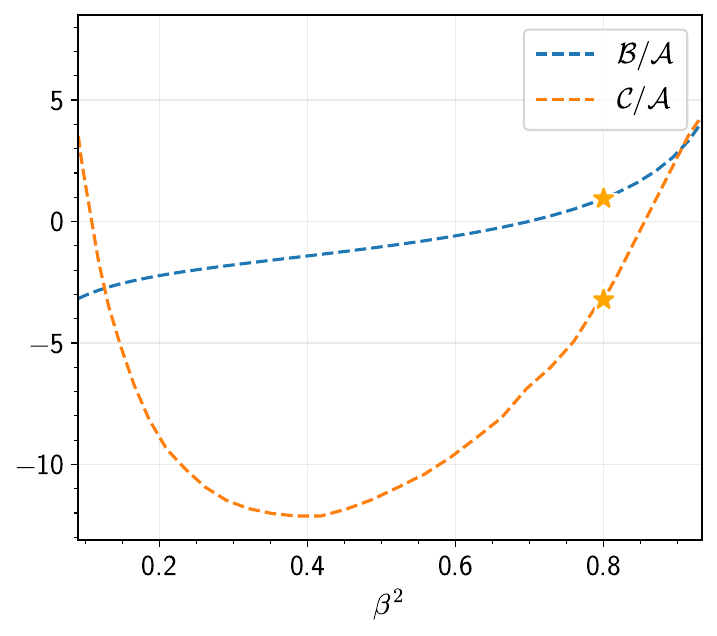}
  \includegraphics[width=0.49\textwidth]{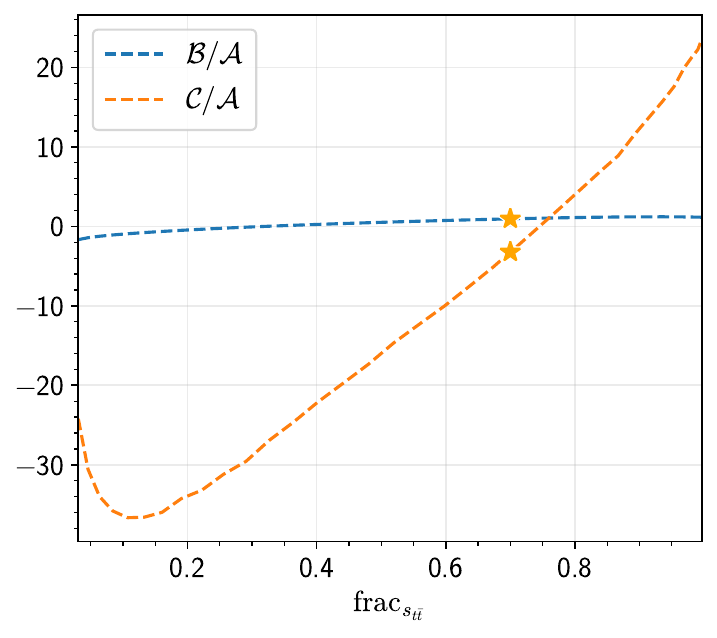}
  \caption{
      One-dimensional slices in $\beta^2$ and $\fstt$ of the one-
      and two-loop amplitudes ${\cal B}$ and ${\cal C}$ as defined
      in~\ref{eq:amp1L} and \ref{eq:amp2L}, normalised to the Born
      amplitude squared ${\cal A}$ from \ref{eq:interference0},
      around the centre point of \ref{eq:centre}.
      The centre point is marked with a star.
      Each plot is an interpolation from around 30 data points.
  }
  \label{fig:slice-beta2-BC_over_A}
\end{figure}

In \ref{fig:slice-beta2-BC_over_A} we have plotted one-dimensional
slices of both ${\cal C}$ and ${\cal B}$ in $\beta^2$ and $\fstt$.
The plots illustrate the difference in behaviour of the one- and
two-loop amplitudes across the parameter space; the two-loop
amplitude changes more rapidly, and is on average more negative.

\begin{figure}
  \centering
  \includegraphics[width=0.49\textwidth]{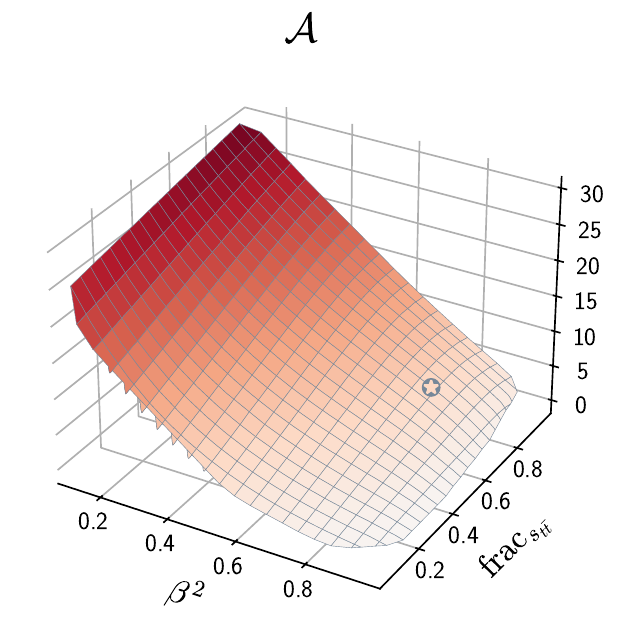}
  \includegraphics[width=0.49\textwidth]{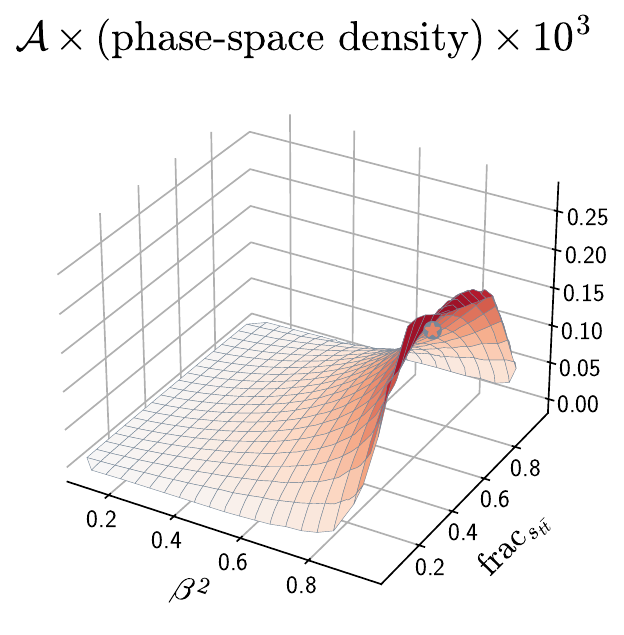}
  \caption{
      Slice of the LO amplitude around the centre point of
      \ref{eq:centre} in $\beta^2$ and $\fstt$.
      On the right the amplitude is multiplied by the phase-space
      density of \ref{eq:PSdensity}.
      The centre point is marked with a star.
  }
  \label{fig:slice-beta2-fstt-A}
\end{figure}

\begin{figure}
  \centering
  \includegraphics[width=0.49\textwidth]{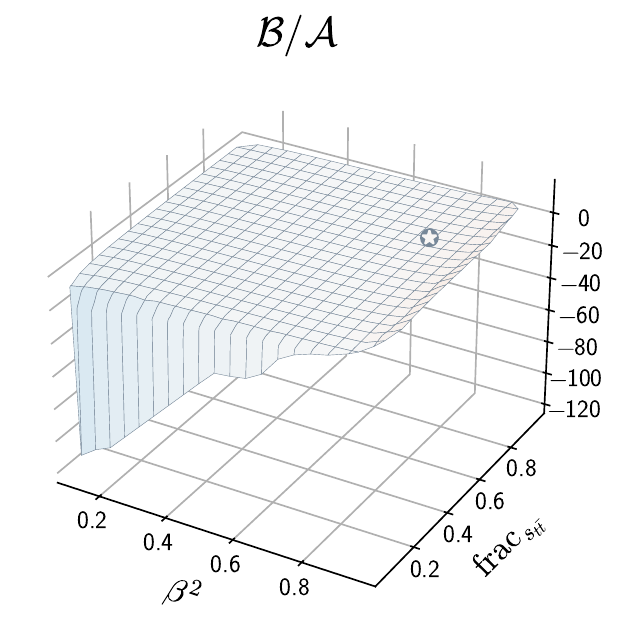}
  \includegraphics[width=0.49\textwidth]{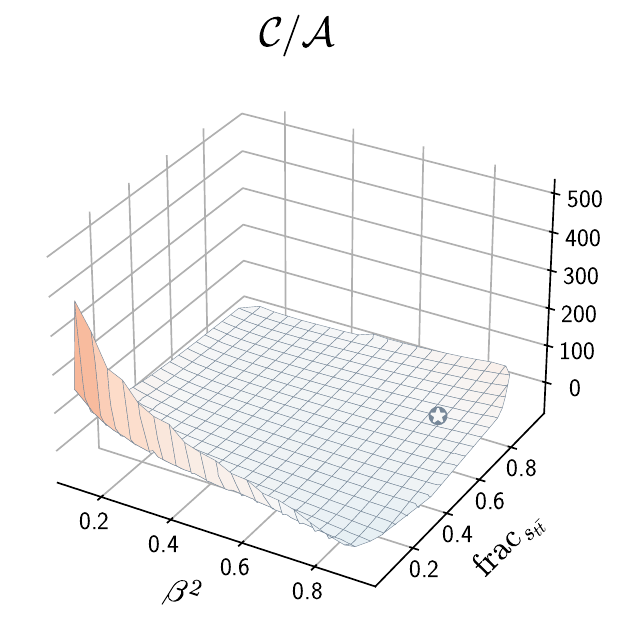}
  \caption{
      Slices of the normalised one-loop (left) and two-loop (right) virtual amplitudes around the centre point of
      \ref{eq:centre} in $\beta^2$ and $\fstt$.
      The centre point is marked with a star.
      Each plot is a linear interpolation of grid of around 500
      data points in total.
  }
  \label{fig:slice-beta2-fstt-BC_over_A}
\end{figure}

\begin{figure}
  \centering
  \includegraphics[width=0.49\textwidth]{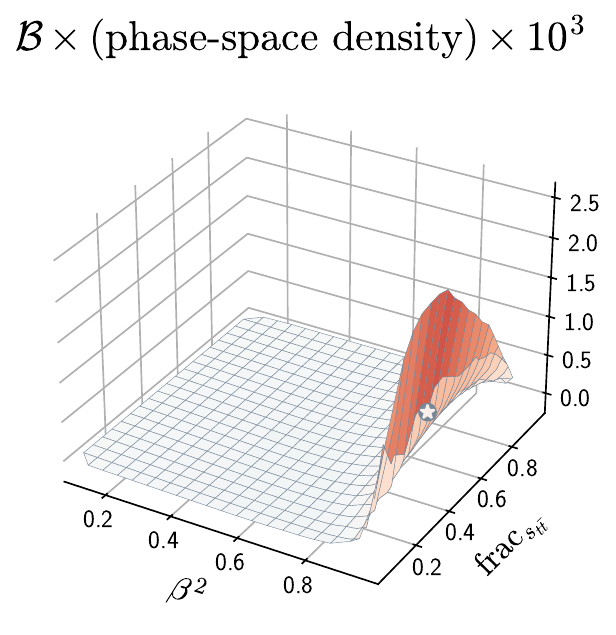}
  \includegraphics[width=0.49\textwidth]{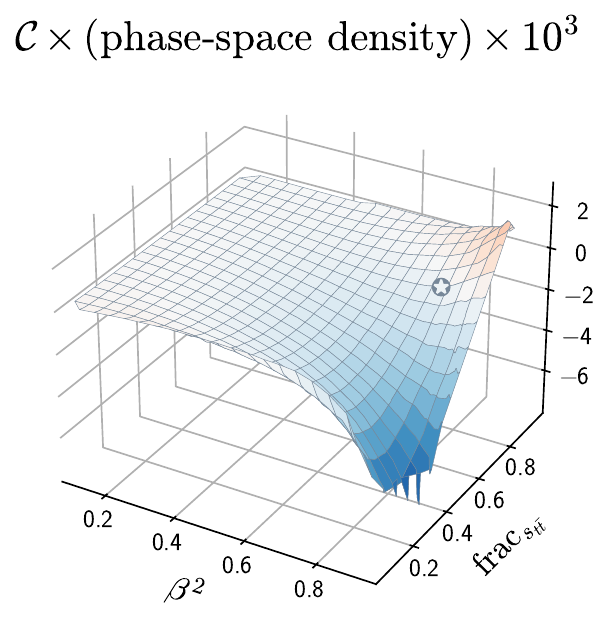}
  \caption{
      Slices of the one-loop (left) and two-loop (right) virtual
      amplitudes multiplied by the phase-space density of
      \ref{eq:PSdensity}, around the centre point of \ref{eq:centre}
      in $\beta^2$ and $\fstt$.
      The centre point is marked with a star.
  }
  \label{fig:slice-beta2-fstt-BC_ps}
\end{figure}

Both in \ref{fig:slice-beta2-BC_over_A} and further it is
convenient to use the LO amplitude ${\cal A}$ as a reference;
a slice of ${\cal A}$ in $\beta^2$ and $\fstt$ is presented
in \ref{fig:slice-beta2-fstt-A}.
Note that once the phase-space density factor of \ref{eq:PSdensity}
is included to obtain the event probability density, the regions of
low-$\beta^2$, low-$\fstt$, and high-$\fstt$ are all suppressed.
This suppression is important because starting at the one-loop
level the amplitude develops a Coulomb-type singularity in the
low-$\fstt$ region.
This singularity can be seen on the slices of ${\cal
B}$ and ${\cal C}$ in $\beta^2$ and $\fstt$ depicted in
\ref{fig:slice-beta2-fstt-BC_over_A}.
The inclusion of the phase-space density however suppresses this
divergence, as can be observed in \ref{fig:slice-beta2-fstt-BC_ps}.

\begin{figure}
  \centering
  \includegraphics[width=0.49\textwidth]{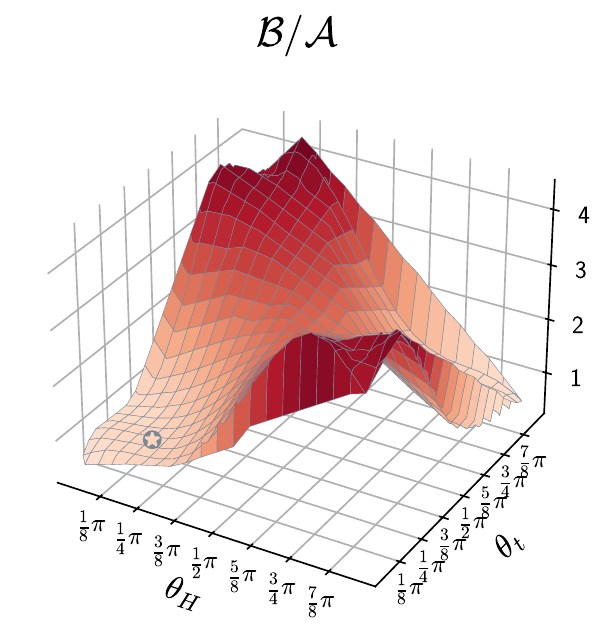}
  \includegraphics[width=0.49\textwidth]{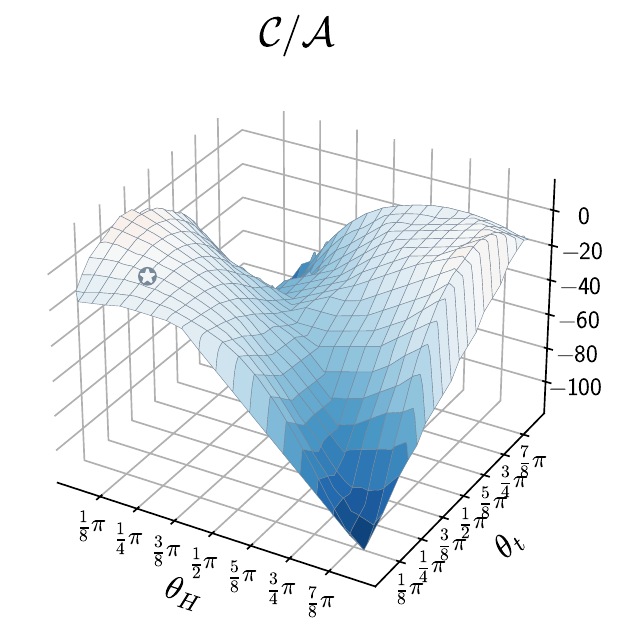}
  \caption{
      Slice of the normalised one-loop (left) and two-loop
      (right) virtual amplitudes around the centre point of
      \ref{eq:centre} in $\theta_H$ and $\theta_t$.
      The centre point is marked with a star.
  }
  \label{fig:slice-theta_h-theta_t-BC_over_A}
\end{figure}

\begin{figure}
  \centering
  \includegraphics[width=0.49\textwidth]{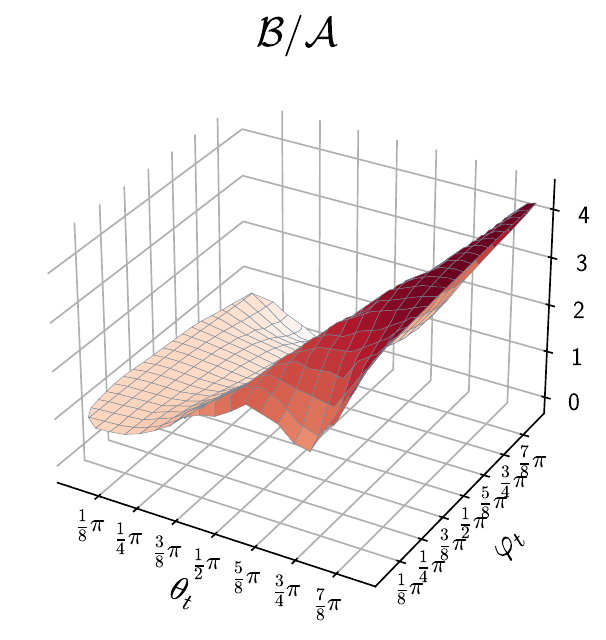}
  \includegraphics[width=0.49\textwidth]{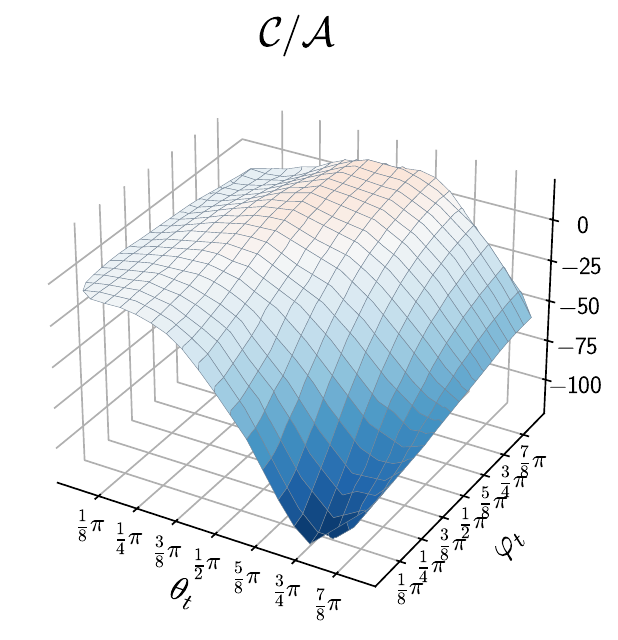}
  \caption{
      Slices of the normalised one-loop (left) and two-loop (right) virtual amplitudes around the centre point of
      \ref{eq:centre} in $\theta_t$ and $\varphi_t$.
      The centre point is marked with a star.
  }
  \label{fig:slice-theta_t-phi_t-BC_over_A}
\end{figure}

To further illustrate the difference in behaviour between the one-
and the two-loop level results, we present the slice in $\theta_H$
and $\theta_t$ in \ref{fig:slice-theta_h-theta_t-BC_over_A}.
A similar slice in $\theta_t$ and $\varphi_t$ is presented in
\ref{fig:slice-theta_t-phi_t-BC_over_A}.

\begin{figure}
  \centering
  \includegraphics[width=0.24\textwidth]{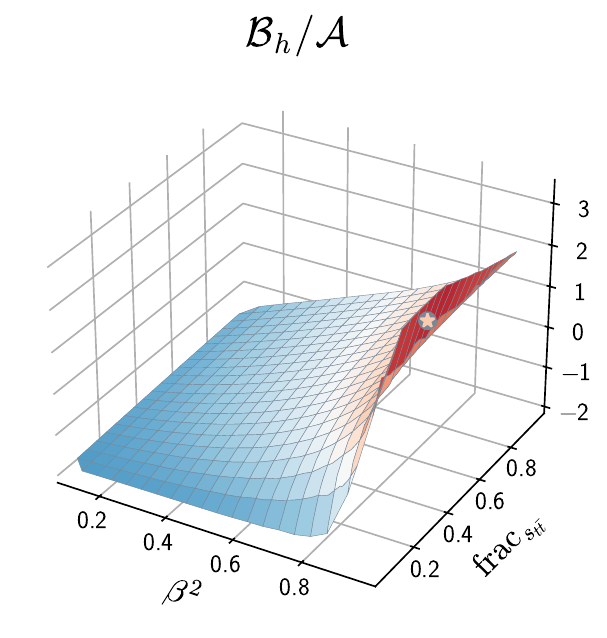}
  \includegraphics[width=0.24\textwidth]{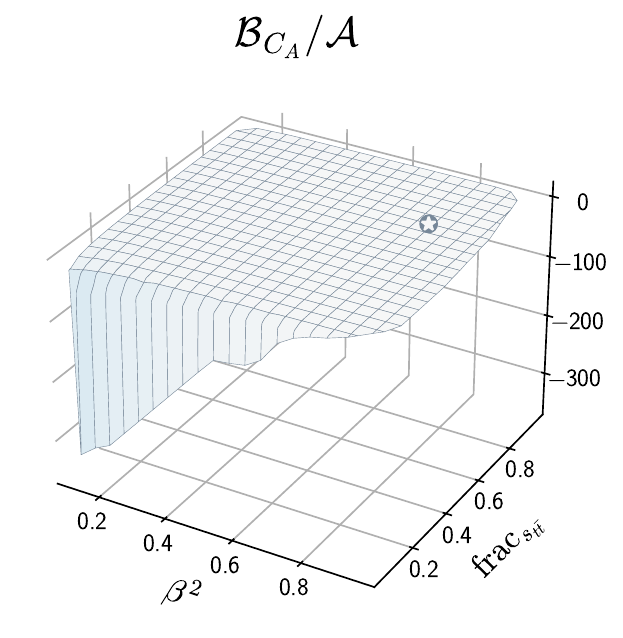}
  \includegraphics[width=0.24\textwidth]{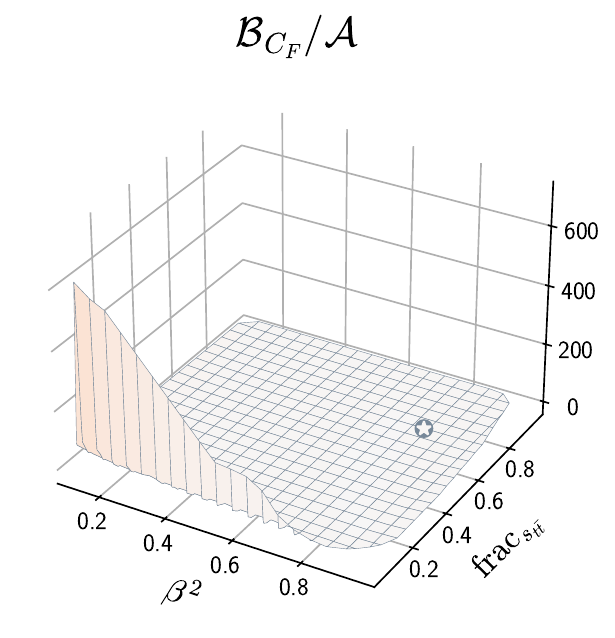}
  \includegraphics[width=0.24\textwidth]{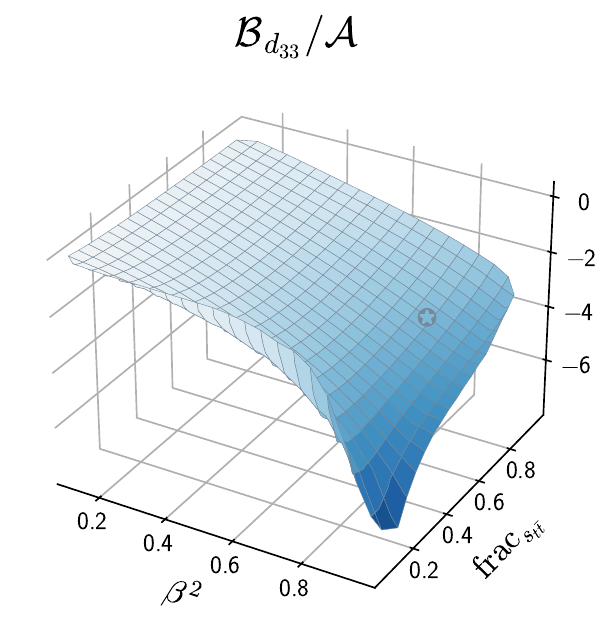}
  \includegraphics[width=0.24\textwidth]{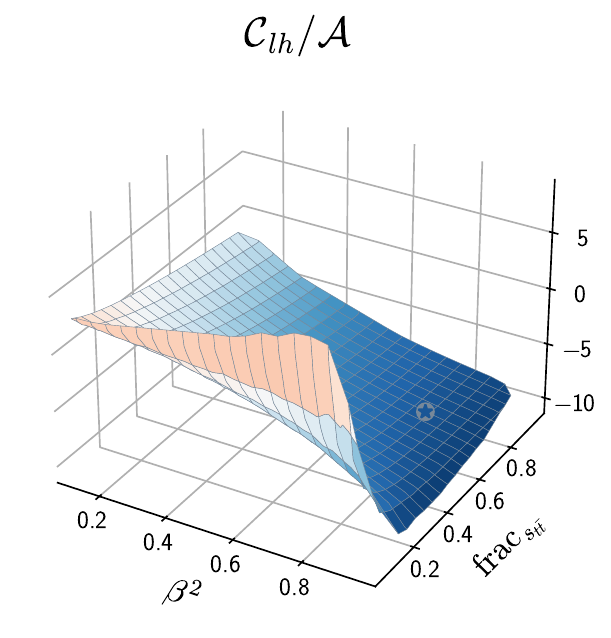}
  \includegraphics[width=0.24\textwidth]{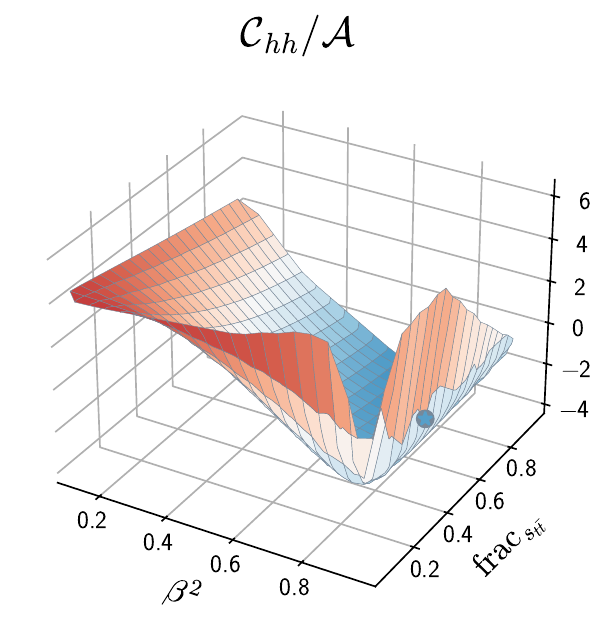}
  \includegraphics[width=0.24\textwidth]{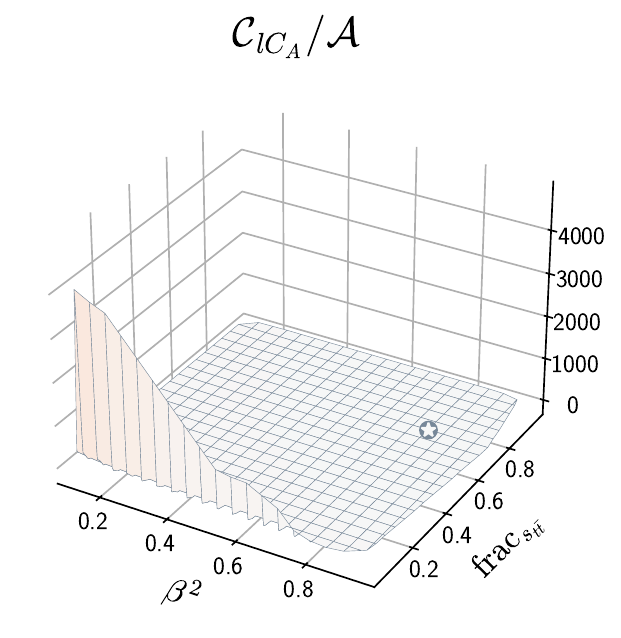}
  \includegraphics[width=0.24\textwidth]{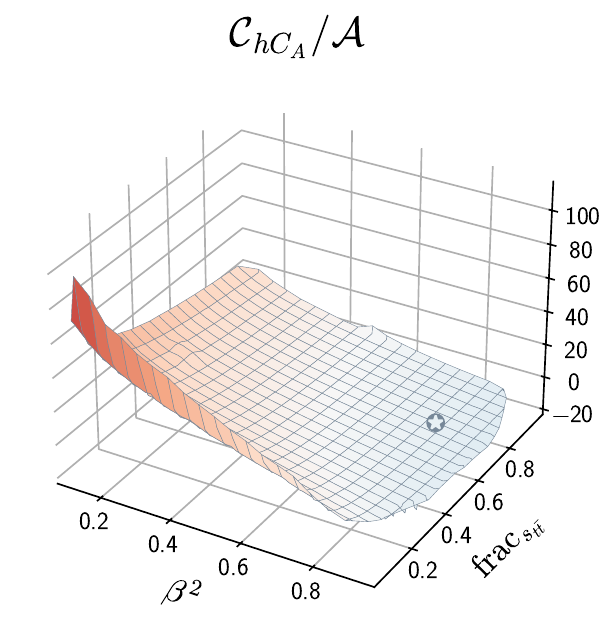}
  \includegraphics[width=0.24\textwidth]{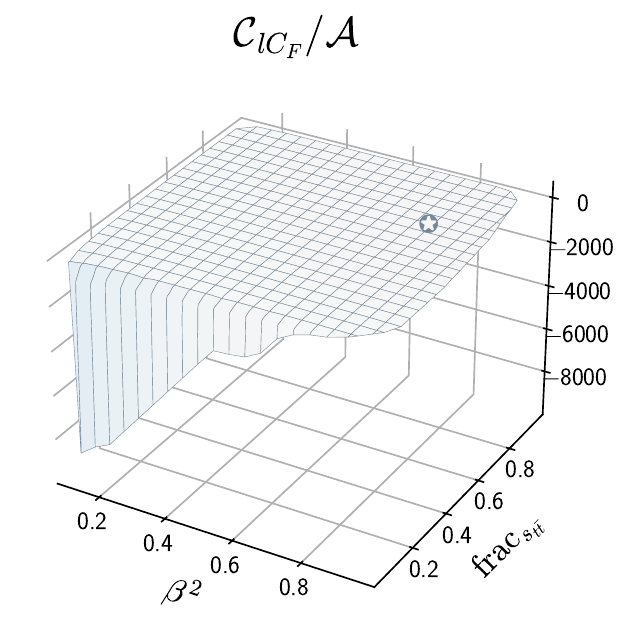}
  \includegraphics[width=0.24\textwidth]{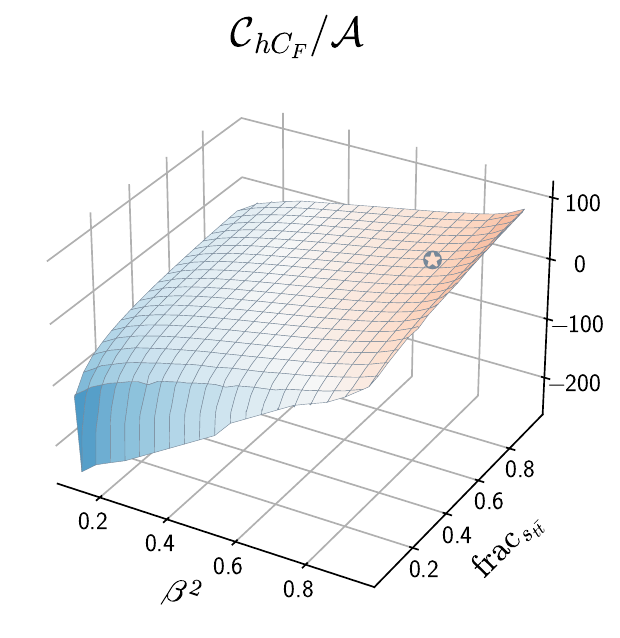}
  \includegraphics[width=0.24\textwidth]{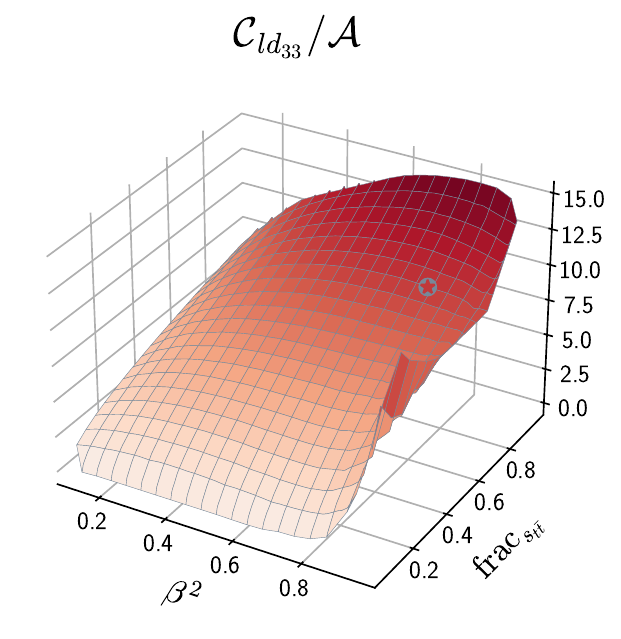}
  \includegraphics[width=0.24\textwidth]{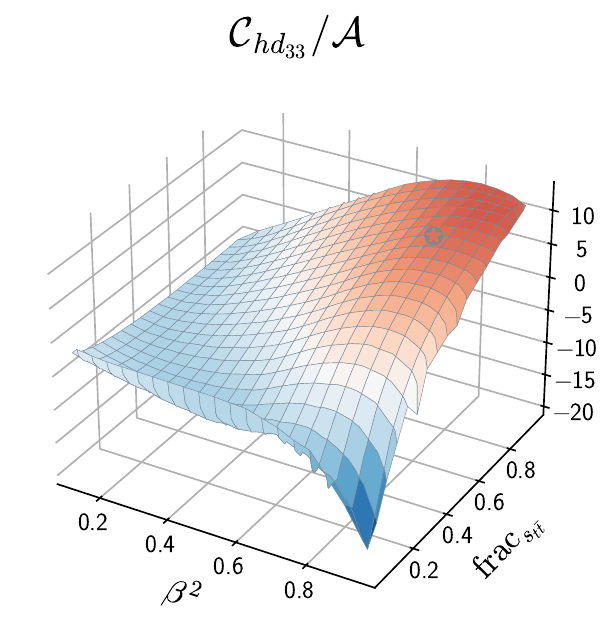}
  \caption{
      Contributions from the individual colour factors to the one-
      and two-loop amplitudes for phase-space slices around the
      centre point of \ref{eq:centre} in $\beta^2$ and $\fstt$.
      The centre point is marked with a star.
  }
  \label{fig:slice-beta2-fstt-BCxx_over_A}
\end{figure}

Finally, we illustrate the difference in behaviour between
different components of ${\cal B}$ and ${\cal C}$ in
\ref{fig:slice-beta2-fstt-BCxx_over_A}, with a slice in $\beta^2$
and $\fstt$ for each of the individual components, aside from
${\cal B}_{l}$, ${\cal C}_{ll}$, which are not plotted because
their ratio to ${\cal A}$ is constant.

%% file: appendix.tex
\subsection{Details of UV renormalisation}
\label{app:UV}
The renormalisation constants expanded up to order $\as^2$ and $\eps^2$ are given by~\cite{Baernreuther:2013caa}:
\begin{align}
 Z_q &= 1 + \left(\frac{\as}{2\pi}\right)^2 C_F T_F n_h\,\delta_{Z_q,n_h}^{(2)}\;,\nn\\
 Z_Q &= 1+ \left(\frac{\as}{2\pi}\right) C_F\,\delta_{Z_Q}^{(1)}  \nn\\
     & +\left(\frac{\as}{2\pi}\right)^2C_F\Big\{T_F n_h \,\delta_{Z_Q,n_h}^{(2)}+T_F n_l \,\delta_{Z_Q,n_l}^{(2)}+C_F  \,\delta_{Z_Q,C_F}^{(2)}+C_A  \,\delta_{Z_Q,C_A}^{(2)}\Big\}\;,\nn\\
 Z_m &= 1+ \left(\frac{\as}{2\pi}\right) C_F\,\delta_{Z_m}^{(1)}  \nn\\
     & +\left(\frac{\as}{2\pi}\right)^2C_F\Big\{T_F n_h
     \,\delta_{Z_m,n_h}^{(2)}+T_F n_l \,\delta_{Z_m,n_l}^{(2)}+C_F
     \,\delta_{Z_m,C_F}^{(2)}+C_A
     \,\delta_{Z_m,C_A}^{(2)}\Big\}\;,
   \end{align}
  \begin{align}
  Z_\alpha&=1- \left(\frac{\as}{2\pi}\right)\frac{\beta_0}{2 \eps}
   +  \left( \frac{\as}{2\pi}\right)^2 \left(
     \frac{\beta_0^2}{4\eps^2}
     - \frac{\beta_1}{8 \eps}
     \right)\nn\\
     &=1+ \left(\frac{\as}{2\pi}\right)
      \left(
       C_A \delta_{\as,C_A}^{(1)}
       + T_F \left(n_l + n_h\right) \delta_{\as,N_f}^{(1)}
      \right)\nn\\
   &+  \left( \frac{\as}{2\pi}\right)^2
      \left(
       C_A^2 \delta_{\as,C_A^2}^{(2)}
       + C_A T_F  \left(n_l + n_h\right) \delta_{\as,C_A N_f}^{(2)}
       + C_F T_F  \left(n_l + n_h\right) \delta_{\as,C_F N_f}^{(2)}
      \right.\nn\\
   & \left.\qquad
       + T_F^2 \left(n_l^2 + 2n_l n_h + n_h^2\right) \delta_{\as,N_f^2}^{(2)}
      \right)
\;,\nn\\
\beta_0 &= \frac{11}{3} C_A- \frac{4}{3}T_F N_f\;,\;
          \beta_1 = \frac{34}{3} C_A^2 -  \frac{20}{3} C_AT_F N_f - 4 C_F T_F N_f \;,  \nn\\
 \zeta_{\as} &= 1+ \left(\frac{\as^{(n_l)}}{2\pi}\right) T_F n_h\,\delta_{ \zeta_{\as}}^{(1)}  + \left(\frac{\as^{(n_l)}}{2\pi}\right)^2 T_F n_h\,\delta_{ \zeta_{\as}}^{(2)}  \; .
 \label{eq:renormZs}
\end{align}
The renormalised amplitude in powers of $\aslpi$, up to the desired accuracy, is given by
\begin{align}
  & {\cal A}^R\left(\asl,y_t,m,\mu,\eps\right)
  =
   \frac{4 \pi \asl y_t}{\sqrt{2}m} \Bigg\{{\cal A}_{0}^{R}\left(\eps\right)
\nn\\
  & \quad
   + \left(\aslpi \right) {\cal A}_{1}^{R}\left(\frac{\mu}{m},\eps\right)
   + \left(\aslpi \right)^2 {\cal A}_{2}^{R}\left(\frac{\mu}{m},\eps\right)
   + {\cal O}\left({\asl}^3 \right) \Bigg\}\;.  \label{eq:renorm_expansion}
\end{align}
The explicit coefficients of the expansion are given by
\begin{align}
  {\cal A}_{0}^{R}(\eps) =&  {\cal A}_{0}(\eps)
  \nn\\
  {\cal A}_{1}^{R}\left(\frac{\mu}{m},\eps\right) =&  \left(Z_Q^{(1)}+Z_{\as}^{(1)}+Z_m^{(1)}+ \zeta_{\as} ^{(1)}\right){\cal A}_{0}(\eps)+Z_m^{(1)} {\cal A}_{0}^{(1)mct}(\eps)+\left(\frac{\mu^2}{m^2}\right)^\eps S_\eps^{-1} {\cal A}_{1}(\eps)
   \nn\\
  {\cal A}_{2}^{R}\left(\frac{\mu}{m},\eps\right) =& \left(Z_q^{(2)}+Z_Q^{(2)}+Z_{\as}^{(2)}+Z_m^{(2)}+Z_Q^{(1)}Z_{\as}^{(1)}+Z_Q^{(1)}Z_m^{(1)}+Z_{\as}^{(1)}Z_m^{(1)}
  \right. \nn\\
  &\left.\quad
  +\zeta_{\as} ^{(2)}+2\zeta_{\as} ^{(1)}\left(Z_Q^{(1)}+Z_{\as}^{(1)}+Z_m^{(1)}\right)\right){\cal A}_{0}(\eps)
\nn\\
  &
  +\frac{1}{2}(Z_m^{(1)})^2 {\cal A}_{0}^{(2)mct}(\eps)+\left(Z_m^{(2)}+Z_m^{(1)}\left(Z_Q^{(1)}+Z_{\as}^{(1)}+Z_m^{(1)}+2\zeta_{\as} ^{(1)}\right)\right) {\cal A}_{0}^{(1)mct}(\eps)
\nn\\
  &
  +\left(\frac{\mu^2}{m^2}\right)^\eps S_\eps^{-1} \left(\left(Z_Q^{(1)}+2Z_{\as}^{(1)}+Z_m^{(1)}+ 2\zeta_{\as} ^{(1)}\right){\cal A}_{1}(\eps)+Z_m^{(1)} {\cal A}_{1}^{(1)mct}(\eps)\right)
\nn\\
  &
   + \left(\frac{\mu^2}{m^2}\right)^{2\eps} S_\eps^{-2} {\cal A}_{2}(\eps)\;. \label{eq:renorm_coeff}
\end{align}
Below we list the renormalisation constants entering \ref{eq:renormZs} up to order $\eps^2$.

\begin{align}
\delta_{Z_q,n_h}^{(2)}&=
  \frac{1}{4\eps}  + \frac{1}{2} \lmu{}  - \frac{5}{24}\;,\nn\\
\delta_{Z_Q}^{(1)}&= C_F\Big\{
- \frac{3}{2\eps}
- 2
- \frac{3}{2}\lmu{}
- 4 \eps
- 2 \eps \lmu{}
- \frac{3}{4} \eps \lmu^2
- \frac{\pi^2}{8} \eps
- 8 \eps^2
- 4 \eps^2 \lmu{}
- \eps^2 \lmu^2
\nn\\
&
- \frac{1}{4} \eps^2 \lmu^3
- \frac{\pi^2}{6} \eps^2
- \frac{\pi^2}{8} \eps^2 \lmu{}
+ \frac{1}{2} \eps^2 \zeta_3
        \Big\}\;,\nn\\
  \delta_{Z_Q,n_h}^{(2)}&=
\frac{1}{4\eps}
+ \frac{1}{\eps} \lmu{}
+ \frac{947}{72}
+ \frac{11}{6} \lmu{}
+ \frac{3}{2} \lmu^2
 - \frac{5\pi^2}{4}\;,\nn\\
\delta_{Z_Q,n_l}^{(2)}&=
- \frac{1}{2\eps^2}
+ \frac{11}{12\eps}
+ \frac{113}{24}
+ \frac{19}{6} \lmu{}
+ \frac{1}{2} \lmu^2
+ \frac{\pi^2}{3}\;,\nn\\
\delta_{Z_Q,C_F}^{(2)}&=
\frac{9}{8\eps^2}
+ \frac{51}{16\eps}
+ \frac{9}{4\eps} \lmu{}
+ \frac{433}{32}
+ \frac{51}{8} \lmu{}
+ \frac{9}{4} \lmu^2
- \frac{49\pi^2}{16}
+ 4 \ln(2\pi^2)
- 6 \zeta_3\;,\nn\\
 \delta_{Z_Q,C_A}^{(2)}&=
\frac{11}{8\eps^2}
- \frac{127}{48\eps}
- \frac{1705}{96}
- \frac{215}{24} \lmu{}
- \frac{11}{8} \lmu^2
+ \frac{5\pi^2}{4}
- 2 \ln(2\pi^2)
+3 \zeta_3
\end{align}
\begin{align}
  \delta_{Z_m}^{(1)} &=
- \frac{3}{2\eps}
- 2
- \frac{3}{2}\lmu{}
- 4 \eps
- 2 \eps \lmu{}
- \frac{3}{4} \eps \lmu^2
- \frac{\pi^2}{8} \eps
- 8 \eps^2
- 4 \eps^2 \lmu{}
- \eps^2 \lmu^2\nn\\
& - \frac{1}{4} \eps^2 \lmu^3
- \frac{\pi^2}{6} \eps^2
- \frac{\pi^2}{8} \eps^2 \lmu{}
                       + \frac{1}{2} \eps^2 \zeta_3\;,\nn\\
 \delta_{Z_m,n_h}^{(2)} &=
- \frac{1}{2\eps^2}
+ \frac{5}{12\eps}
+ \frac{143}{24}
+ \frac{13}{6} \lmu{}
+ \frac{1}{2} \lmu^2
- \frac{2\pi^2}{3}\;,\nn\\
 \delta_{Z_m,n_l}^{(2)} &=
- \frac{1}{2\eps^2}
+ \frac{5}{12\eps}
+ \frac{71}{24}
+ \frac{13}{6} \lmu{}
+ \frac{1}{2} \lmu^2
+ \frac{\pi^2}{3}\;,\nn\\
\delta_{Z_m,C_F}^{(2)} &=
\frac{9}{8\eps^2}
+ \frac{45}{16\eps}
+ \frac{9}{4\eps} \lmu{}
+ \frac{199}{32}
+ \frac{45}{8} \lmu{}
+ \frac{9}{4} \lmu^2
- \frac{17\pi^2}{16}
+ 2 \ln(2\pi^2)
- 3 \zeta_3\;,\nn\\
  \delta_{Z_m,C_A}^{(2)} &=
\frac{11}{8\eps^2}
- \frac{97}{48\eps}
- \frac{1111}{96}
- \frac{185}{24} \lmu{}
- \frac{11}{8} \lmu^2
+ \frac{\pi^2}{3}
- \ln(2\pi^2)
+ \frac{3}{2} \zeta_3\;,\nn\\
&\begin{aligned}
  \delta_{\as,C_A}^{(1)} &=
-\frac{11}{6\eps}\;,
&  \delta_{\as,N_f}^{(1)} &=
\frac{2}{3\eps}\;,\\
  \delta_{\as,C_A^2}^{(2)} &=
\frac{121}{36\eps^2}
- \frac{17}{12\eps}\;,
& \delta_{\as,C_A N_f}^{(2)} &=
- \frac{22}{9\eps^2}
+ \frac{5}{6\eps}\;,\\
  \delta_{\as,C_F N_f}^{(2)} &=
\frac{1}{2\eps}\;,
& \delta_{\as,N_f^2}^{(2)} &=
\frac{4}{9\eps^2}\;,
\end{aligned}
\end{align}
where $\lmu{} = \ln \mu^2/m^2$.
The decoupling coefficients are given  by~\cite{Bernreuther:1981sg,Baernreuther:2013caa}
\begin{align}
\delta_{\zeta_{\as}}^{(1)} &=
\frac{2}{3} \lmu{}
+ \frac{1}{3} \eps \lmu^2
+ \frac{\pi^2}{18} \eps
+ \frac{1}{9} \eps^2 \lmu^3
+ \frac{\pi^2}{18} \eps^2 \lmu{}
- \frac{2}{9} \eps^2 \zeta_3\;,\\
\delta_{\zeta_{\as}}^{(2)} &=
\frac{4}{9} T_F n_h
\lmu^2
+ C_F\biggl[
\frac{15}{4}
+ \lmu{}
\biggr]
+ C_A \biggl[
- \frac{8}{9}
+ \frac{5}{3} \lmu{}
\biggr]\;.
\end{align}

\noindent Finally, we list the explicit expressions for the components ${\cal A}$, ${\cal B}_i$ and ${\cal C}_i$ in \ref{eq:interference2} and \ref{eq:interference1}.\footnote{Non calligraphic $A$, $B_i$ and $C_i$ denote the coefficients of the decomposition of the bare amplitude interference.}
\begin{align}
{\cal A}=&\,A
\nn\\
{\cal B}_{l}=&\delta_{Z_{\as,N_f}}^{(1)}\,A+\left(\frac{\mu^2}{m^2}\right)^\eps S_\eps^{-1}\,B_{l}
\nn\\
{\cal B}_{h}=&\left(\delta_{Z_{\as,N_f}}^{(1)}+\delta_{\zeta_{\as}}^{(1)}\right) A+\left(\frac{\mu^2}{m^2}\right)^\eps S_\eps^{-1}\,B_{h}
\nn\\
{\cal B}_{C_F}=&\left(\delta_{Z_{Q}}^{(1)}+\delta_{Z_{m}}^{(1)}\right) A+\delta_{Z_{m}}^{(1)}\,A^{(1)mct}+\left(\frac{\mu^2}{m^2}\right)^\eps S_\eps^{-1}\,B_{C_F}
\nn\\
{\cal B}_{C_A}=&\delta_{Z_{\as,C_A}}^{(1)}\,A+\left(\frac{\mu^2}{m^2}\right)^\eps S_\eps^{-1}\,B_{C_A}
\nn\\
{\cal B}_{d_{33}}=&\left(\frac{\mu^2}{m^2}\right)^\eps S_\eps^{-1}\,B_{d_{33}}
                    \label{eq:calA}
\end{align}
\begin{align}
{\cal C}_{ll}=&\delta_{Z_{\as,N_f^2}}^{(2)}\,A
+\left(\frac{\mu^2}{m^2}\right)^\eps S_\eps^{-1}\left[2\delta_{Z_{\as,N_f}}^{(1)}\,B_{l}\right]
+\left(\frac{\mu^2}{m^2}\right)^{2\eps} S_\eps^{-2}\,C_{ll}
\nn\\
{\cal C}_{lh}=&2\left(\delta_{Z_{\as,N_f^2}}^{(2)}+\delta_{Z_{\as,N_f}}^{(1)}\delta_{\zeta_{\as}}^{(1)}\right) A
+\left(\frac{\mu^2}{m^2}\right)^\eps S_\eps^{-1}\left[2\left(\delta_{Z_{\as,N_f}}^{(1)}+\delta_{\zeta_{\as}}^{(1)}\right) B_{l}+2\delta_{Z_{\as,N_f}}^{(1)}\,B_{h}\right]
\nn\\&
+\left(\frac{\mu^2}{m^2}\right)^{2\eps} S_\eps^{-2}\,C_{lh}
\nn\\
{\cal C}_{hh}=&\left(\delta_{Z_{\as,N_f^2}}^{(2)}+2\delta_{Z_{\as,N_f}}^{(1)}\delta_{\zeta_{\as}}^{(1)}+\delta_{\zeta_{\as,n_h}}^{(2)}\right) A
+\left(\frac{\mu^2}{m^2}\right)^\eps S_\eps^{-1}\left[2\left(\delta_{Z_{\as,N_f}}^{(1)}+\delta_{\zeta_{\as}}^{(1)}\right) B_{h}\right]
\nn\\&
+\left(\frac{\mu^2}{m^2}\right)^{2\eps} S_\eps^{-2}\,C_{hh}
  \label{eq:calC1}
\end{align}
\begin{align}
{\cal C}_{lC_F}=&
\left(\delta_{Z_{Q,n_l}}^{(2)}+\delta_{Z_{m,n_l}}^{(2)}+\delta_{Z_{\as,C_F N_f}}^{(2)}+\delta_{Z_{Q}}^{(1)}\delta_{Z_{\as,N_f}}^{(1)}+\delta_{Z_{m}}^{(1)}\delta_{Z_{\as,N_f}}^{(1)}\right) A
\nn\\&
+\left(\delta_{Z_{\as,N_f}}^{(1)}\delta_{Z_{m}}^{(1)}+\delta_{Z_{m,n_l}}^{(2)}\right) A^{(1)mct}
\nn\\&
+\left(\frac{\mu^2}{m^2}\right)^\eps S_\eps^{-1}\left[2\delta_{Z_{\as,N_f}}^{(1)}\,B_{C_F}+\left(\delta_{Z_{Q}}^{(1)}+\delta_{Z_{m}}^{(1)}\right) B_{l}+\delta_{Z_{m}}^{(1)}\,B_{l}^{(1)mct}\right]
\nn\\&
+\left(\frac{\mu^2}{m^2}\right)^{2\eps} S_\eps^{-2}\,C_{lC_F}
\nn\\
{\cal C}_{lC_A}=&
\delta_{Z_{\as,C_A N_f}}^{(2)}\,A
\nn\\&
+\left(\frac{\mu^2}{m^2}\right)^\eps S_\eps^{-1}\left[2\delta_{Z_{\as,N_f}}^{(1)}\,B_{C_A}+2\delta_{Z_{\as,C_A}}^{(1)}\,B_{l}\right]
\nn\\&
+\left(\frac{\mu^2}{m^2}\right)^{2\eps} S_\eps^{-2}\,C_{lC_A}
\nn\\
{\cal C}_{ld_{33}}=&
\left(\frac{\mu^2}{m^2}\right)^\eps S_\eps^{-1}\left[2\delta_{Z_{\as,N_f}}^{(1)}\,B_{d_{33}}\right]
\nn\\&
+\left(\frac{\mu^2}{m^2}\right)^{2\eps} S_\eps^{-2}\,C_{ld_{33}}
\label{eq:calC2}
\end{align}
\begin{align}
{\cal C}_{hC_F}=&
\left(\delta_{Z_{q,n_h}}^{(2)}+\delta_{Z_{Q,n_h}}^{(2)}+\delta_{Z_{m,n_h}}^{(2)}+\delta_{Z_{\as,C_F N_f}}^{(2)}
% +\delta_{Z_{Q}}^{(1)}\delta_{Z_{\as,N_f}}^{(1)}+\delta_{Z_{m}}^{(1)}\delta_{Z_{\as,N_f}}^{(1)}
+\left(\delta_{Z_{\as,N_f}}^{(1)}+2\delta_{\zeta_{\as}}^{(1)}\right)\left(\delta_{Z_{Q}}^{(1)}+\delta_{Z_{m}}^{(1)}\right)+\delta_{\zeta_{\as,C_F}}^{(2)}
\right) A
\nn\\&
+\left(\delta_{Z_{\as,N_f}}^{(1)}\delta_{Z_{m}}^{(1)}+\delta_{Z_{m,n_h}}^{(2)}+2\delta_{\zeta_{\as}}^{(1)}\delta_{Z_{m}}^{(1)}\right) A^{(1)mct}
\nn\\&
+\left(\frac{\mu^2}{m^2}\right)^\eps S_\eps^{-1}\left[2\left(\delta_{Z_{\as,N_f}}^{(1)}+\delta_{\zeta_{\as}}^{(1)}\right) B_{C_F}+\left(\delta_{Z_{Q}}^{(1)}+\delta_{Z_{m}}^{(1)}\right) B_{h}+\delta_{Z_{m}}^{(1)}\,B_{h}^{(1)mct}\right]
\nn\\&
+\left(\frac{\mu^2}{m^2}\right)^{2\eps} S_\eps^{-2}\,C_{hC_F}
  \nn
\end{align}
\begin{align}
{\cal C}_{hC_A}=&
\left(\delta_{Z_{\as,C_A N_f}}^{(2)}+\delta_{\zeta_{\as,C_A}}^{(2)}+2\delta_{Z_{\as,C_A}}^{(1)}\delta_{\zeta_{\as}}^{(1)}\right) A
\nn\\&
+\left(\frac{\mu^2}{m^2}\right)^\eps S_\eps^{-1}\left[2\left(\delta_{Z_{\as,N_f}}^{(1)}+\delta_{\zeta_{\as}}^{(1)}\right) B_{C_A}+2\delta_{Z_{\as,C_A}}^{(1)}\,B_{h}\right]
\nn\\&
+\left(\frac{\mu^2}{m^2}\right)^{2\eps} S_\eps^{-2}\,C_{hC_A}
\nn\\
{\cal C}_{hd_{33}}=&
\left(\frac{\mu^2}{m^2}\right)^\eps S_\eps^{-1}\left[2\left(\delta_{Z_{\as,N_f}}^{(1)}+\delta_{\zeta_{\as}}^{(1)}\right) B_{d_{33}}\right]
\nn\\&
+\left(\frac{\mu^2}{m^2}\right)^{2\eps} S_\eps^{-2}\,C_{hd_{33}}
  \label{eq:calC3}
\end{align}

\subsection{Colour basis for \texorpdfstring{$q_{i_1}\bar{q}_{i_2}\to t_{f_1}\bar{t}_{f_2}H$}{qq→ttH} }
\label{app:Colour}

The construction of the colour basis in this section describes the case of quarks in the defining representation of a $SU(N)$ gauge group for conciseness. 
Nevertheless, the procedure can be extended to include additional orthogonal colour vectors for a more general gauge theory.

Since outgoing quark and incoming antiquark transform under colour representation $N$ and outgoing antiquark and incoming quark transform under colour representation $\bar{N}$, there are only two possibilities to combine colour and anticolour to get a colour conserving scattering amplitude.
Thus, we have a two-dimensional colour space. Possible colour
structures are obtained with
\begin{align}
 \ket{c_1}&=\left(t^a\right)_{i_2i_1}\left(t^a\right)_{f_1f_2}
& \ket{c_2}&=\delta_{i_2i_1}\delta_{f_1f_2}
& \ket{c_3}&=\delta_{i_2f_2}\delta_{f_1i_1}
\end{align}

\begin{align}
   \left(\braket{c_k}{c_l}\right)
  =
\left( \begin{matrix}
   T_FC_FN_C & 0 & C_FN_C  \\
   0 & N_C^2 & N_C  \\
   C_FN_C & N_C & N_C^2
\end{matrix} \right)
\end{align}

\noindent $\ket{c_1}$ and $\ket{c_2}$ are the best choice for the basis, since the vectors are orthogonal. In addition, the tree level is proportional to $\ket{c_1}$, hence we only need to recalculate the amplitude interference with colour projector on $\ket{c_2}$, as explained in the following.

We expand the amplitudes in terms of colour vectors
\begin{align}
   \ket{{\cal A}^{(0)}}
  = &
   \ket{A^{(0)}}\otimes \ket{c_1}
\nn\\
   \ket{{\cal A}^{(n)}}
  = &
   \ket{A^{(n)}_1}\otimes \ket{c_1}
   + \ket{A^{(n)}_2}\otimes \ket{c_2}\;.
\end{align}
Now we calculate the general form for the IR poles
\begin{align}
   \bra{{\cal A}^{(0)}}\mathbf{Z}_i \ket{{\cal A}^{(n)}}
  = &
   \braket{A^{(0)}}{A_1^{(n)}} \cdot \bra{c_1}\mathbf{Z}_i\ket{c_1}
   + \braket{A^{(0)}}{A_2^{(n)}} \cdot \bra{c_1}\mathbf{Z}_i\ket{c_2}\;.
\end{align}

\noindent We obtain the coefficients of the elements of the colour matrix by using the projectors
\begin{align}
   \ket{p_1}
  = &
   \frac{1}{ T_FC_FN_C } \ket{c_1}
   &\ket{p_2}
  = &
   \frac{1}{N_C^2} \ket{c_2}
\end{align}
as colour vector of the tree amplitude
\begin{align}
   \ket{{\cal P}_1}
  = &
   \ket{A^{(0)}}\otimes \ket{p_1}
   &\ket{{\cal P}_2}
  = &
   \ket{A^{(0)}}\otimes \ket{p_2}\;,
\end{align}
and multiply them on the loop amplitude
\begin{align}
   \braket{{\cal P}_1}{{\cal A}^{(n)}}
  = &
   \braket{A^{(0)}}{A_1^{(n)}}\cdot \underbrace{\braket{p_1}{c_1}}_{1}
  =
   \frac{1}{T_FC_FN_C}\braket{{\cal A}^{(0)}}{{\cal A}^{(n)}}\;.
\nn\\
   \braket{{\cal P}_2}{{\cal A}^{(n)}}
  = &
   \braket{A^{(0)}}{A_2^{(n)}}\cdot \underbrace{\braket{p_2}{c_2}}_{1}
\end{align}

\noindent $\bra{c_k}\mathbf{Z}\ket{c_l}$ is obtained by evaluating $\left(\bra{c_k}\mathbf{T}_{I}\cdot \mathbf{T}_{J}\ket{c_l}\right)$
and $\left(\bra{c_k}i f^{abc}\mathbf{T}^a_{I}\mathbf{T}^b_{J}\mathbf{T}^c_{K}\ket{c_l}\right)$.
\begin{align}
   \left(\bra{c_k}\mathbf{T}_{1}\cdot \mathbf{T}_{2}\ket{c_l}\right)
  = &
\left(\begin{matrix}
   -T_FC_FN_C\left(C_F-\frac{C_A}{2}\right) & 0   \\
   0 & -C_FN_C^2
\end{matrix}\right)
\nn\\
   \left(\bra{c_k}\mathbf{T}_{1}\cdot \mathbf{T}_{3}\ket{c_l}\right)
  = &
\left(\begin{matrix}
   -d_{33}-\frac{T_FC_FC_AN_C}{4} & -T_FC_FN_C   \\
   -T_FC_FN_C & 0
\end{matrix}\right)
\nn\\
   \left(\bra{c_k}\mathbf{T}_{1}\cdot \mathbf{T}_{4}\ket{c_l}\right)
  = &
\left(\begin{matrix}
   d_{33}-\frac{T_FC_FC_AN_C}{4} & T_FC_FN_C   \\
   T_FC_FN_C & 0
\end{matrix}\right)
\nn\\
   \left(\bra{c_k}\mathbf{T}_{2}\cdot \mathbf{T}_{3}\ket{c_l}\right)
  = &
\left(\begin{matrix}
   d_{33}-\frac{T_FC_FC_AN_C}{4} & T_FC_FN_C   \\
   T_FC_FN_C & 0
\end{matrix}\right)
\nn\\
   \left(\bra{c_k}\mathbf{T}_{2}\cdot \mathbf{T}_{4}\ket{c_l}\right)
  = &
\left(\begin{matrix}
   -d_{33}-\frac{T_FC_FC_AN_C}{4} & -T_FC_FN_C   \\
   -T_FC_FN_C & 0
\end{matrix}\right)
\nn\\
   \left(\bra{c_k}\mathbf{T}_{3}\cdot \mathbf{T}_{4}\ket{c_l}\right)
  = &
\left(\begin{matrix}
   -T_FC_FN_C\left(C_F-\frac{C_A}{2}\right) & 0   \\
   0 & -C_FN_C^2
\end{matrix}\right)
\nn\\
%    \left(\bra{c_k}1\ket{c_l}\right)
%   = &
% \left(\begin{matrix}
%    T_FC_FN_C & 0   \\
%    0 & N_C^2
% \end{matrix}\right)
% \nn\\
   \left(\bra{c_k}i f^{abc}\mathbf{T}^a_{1}\mathbf{T}^b_{3}\mathbf{T}^c_{4}\ket{c_l}\right)
  = &
  \frac{C_A}{2} \left(\begin{matrix}
   0 & -T_FC_FN_C   \\
   T_FC_FN_C & 0
\end{matrix}\right)
\nn\\
   \left(\bra{c_k}i f^{abc}\mathbf{T}^a_{2}\mathbf{T}^b_{3}\mathbf{T}^c_{4}\ket{c_l}\right)
  = &
  \frac{C_A}{2} \left(\begin{matrix}
   0 & T_FC_FN_C   \\
   -T_FC_FN_C & 0
\end{matrix}\right)
\end{align}

\noindent Therefore we need  $\braket{{\cal P}_1}{{\cal A}^{(0)}}$ up to ${\cal O}\left(\eps^2\right)$, and $\braket{{\cal P}_1}{{\cal A}^{(1)}}$ and
$\braket{{\cal P}_2}{{\cal A}^{(1)}}$ up to order ${\cal O}\left(\eps\right)$.

\subsection{Anomalous dimensions}
\label{app:IR}

The anomalous dimensions entering \ref{eq:Gamma}, expanded up to order $\left(\asl\right)^2$,  are given by~\cite{Baernreuther:2013caa}
\begin{align}
   \gamma_q\left(\asl\right)
  = &
   \left(\aslpi\right)C_F\gamma_q^{(1)}
   + \left(\aslpi\right)^2C_F\left\{
    C_A\gamma_{q,C_A}^{(2)}+C_F\gamma_{q,C_F}^{(2)}+T_F n_l\gamma_{q,n_l}^{(2)} \right\},
\nn\\
   \gamma_Q\left(\asl\right)
  = &
   \left(\aslpi\right)C_F\gamma_Q^{(1)}
   + \left(\aslpi\right)^2C_F\left\{
   C_A\gamma_{Q,C_A}^{(2)}+T_F n_l\gamma_{Q,n_l}^{(2)} \right\},
\nn\\
   \gamma_{cusp}\left(\asl\right)
  = &
   \left(\aslpi\right)\gamma_{cusp}^{(1)}
   + \left(\aslpi\right)^2\left\{
   C_A\gamma_{cusp,C_A}^{(2)}+T_F n_l\gamma_{cusp,n_l}^{(2)} \right\},
\nn\\
   \gamma_{cusp}\left(\beta,\asl\right)
  = &
   \gamma_{cusp}\left(\asl\right)\beta\coth\beta
\nn\\
   & + \left(\aslpi\right)^2 2C_A\left\{
   \coth^2\beta\left[
    \text{Li}_3\left(e^{-2\beta}\right)
    +\beta\text{Li}_2\left(e^{-2\beta}\right)
    -\zeta_3+\frac{\pi^2}{6}\beta+\frac{1}{3}\beta^3\right]
   \right.
\nn\\
   &\left. \qquad
   + \coth\beta\left[
    \text{Li}_2\left(e^{-2\beta}\right)
    - 2\beta \log\left(1-e^{-2\beta}\right)
    - \frac{\pi^2}{6}\left(1+\beta\right)
    - \beta^2 - \frac{1}{3}\beta^3\right]
   \right.
\nn\\
   &\left. \qquad
    + \frac{\pi^2}{6} + \zeta_3 + \beta^2
    \right\},
\end{align}
with explicit values
\begin{align}
   \gamma_q^{(1)}
  &=
   - \frac{3}{2},
   &\gamma_Q^{(1)}
  &=
   - 1,
   &\gamma_{cusp}^{(1)}
  &=
    2,
\nn\\
   \gamma_{q,C_A}^{(2)}
  &=
   - \frac{961}{216} - \frac{11\pi^2}{24} + \frac{13}{2}\zeta_3,
   &\gamma_{Q,C_A}^{(2)}
  &=
   - \frac{49}{18} + \frac{\pi^2}{6} - \zeta_3,
   &\gamma_{cusp,C_A}^{(2)}
  &=
   \frac{67}{9} - \frac{\pi^2}{3},
\nn\\
   \gamma_{q,n_l}^{(2)}
  &=
   \frac{65}{54} + \frac{\pi^2}{6},
   &\gamma_{Q,n_l}^{(2)}
  &=
   \frac{10}{9},
   &\gamma_{cusp,n_l}^{(2)}
  &=
   - \frac{20}{9},
\nn\\
   \gamma_{q,C_F}^{(2)}
  &=
   - \frac{3}{8} + \frac{\pi^2}{2} - 6 \zeta_3.
\end{align}

\subsection{Momentum parametrisation}
\label{app:PS}

In the centre-of-mass (COM) frame of the $2\rightarrow 3$ process,
\begin{equation}
q(q_1)+\bar{q}(q_2)\to t(q_t)+\bar{t}(q_{\bar{t}})+H(q_H)\;,
\end{equation}
we have:
\begin{align}
q_{1,2}&=\frac{\sqrt{s}}{2}\left(1,0,0,\pm 1\right),\nonumber \\
q_H&=\left(E_h, p_H \sin \theta_H, 0, p_H \cos \theta_H\right),\nonumber \\
q_{t\bar{t}}&=\left(\sqrt{s_{t\bar{t}}+p_H^2},-p_H \sin \theta_H,0,-p_H \cos \theta_H\right),\nonumber \\
p_H&=\frac{\sqrt{\lambda(s,s_{t\bar{t}},m_H^2)}}{2\sqrt{s}}=\frac{1}{2}\sqrt{s-2\left(s_{t\bar{t}}+m_H^2\right)+\frac{\left(s_{t\bar{t}}-m_H^2\right)^2}{s}},\nn\\
E_H&=\sqrt{p_H^2+m_H^2}.
\end{align}
To describe $q^{\prime}_t$, $q^{\prime} _{\bar{t}}$ in the ``rest frame of $t\bar{t}$ decay",
with the direction of the (space-like) momentum $q_{t\bar{t}}$ as new z-axis in the decay-frame.
The coordinate transformation into the decay-frame is obtained by:
\begin{itemize}
\item[1)] rotation around y-axis with $R_y=\left(\begin{array}{c c c c}
1 & 0 & 0 & 0\\
0 & -\cos\theta_H & 0 & \sin\theta_H\\
0 & 0 & 1 & 0\\
0 & -\sin\theta_H & 0 & -\cos\theta_H
\end{array}\right)$, such that:
\begin{align}
q^{\ast}_{t\bar{t}}=\left(\sqrt{s_{t\bar{t}}+p_H^2},0,0,p_H\right);
\end{align}
\item[2)] boost along new z-axis $B_z=\left(\begin{array}{c c c c}
\frac{\sqrt{s_{t\bar{t}}+p_H^2}}{\sqrt{s_{t\bar{t}}}} & 0 & 0 & -\frac{p_H}{\sqrt{s_{t\bar{t}}}}\\
0 & 1 & 0 & 0\\
0 & 0 & 1 & 0\\
-\frac{p_H}{\sqrt{s_{t\bar{t}}}} & 0 & 0 & \frac{\sqrt{s_{t\bar{t}}+p_H^2}}{\sqrt{s_{t\bar{t}}}}
\end{array}\right)$, such that:
\begin{align*}
q^{\prime} _{t\bar{t}}=\left(\sqrt{s_{t\bar{t}}},0,0,0\right).
\end{align*}
\end{itemize}
\begin{align}
q^{\prime}_t&=\left(\frac{\sqrt{s_{t\bar{t}}}}{2},p^{\prime}_t\sin\theta^{\prime}_t \cos\varphi_t^{\prime},p^{\prime}_t\sin\theta^{\prime}_t \sin\varphi_t^{\prime},p^{\prime}_t\cos\theta^{\prime}_t \right), \nonumber \\
q^{\prime} _{\bar{t}} &=\left(\frac{\sqrt{s_{t\bar{t}}}}{2},-p^{\prime}_t\sin\theta^{\prime}_t \cos\varphi_t^{\prime},-p^{\prime}_t\sin\theta^{\prime}_t \sin\varphi_t^{\prime},-p^{\prime}_t\cos\theta^{\prime}_t \right), \nonumber \\
p^{\prime}_t&=\frac{\sqrt{\lambda(s_{t\bar{t}},m_t^2,m_t^2)}}{2\sqrt{s_{t\bar{t}}}}=\sqrt{\frac{s_{t\bar{t}}}{4}-m_t^2}.
\end{align}

\noindent The transformation back to the COM is given by:
\begin{align}
q_t&=R_y^{-1} B_z^{-1} q^{\prime} _{t}, \nonumber \\
q_{\bar{t}}&=R_y^{-1} B_z^{-1} q^{\prime} _{\bar{t}}\;.
\end{align}

\noindent The most general set of momenta is given by a global rotation around the z-axis in the COM frame.

\subsection{Integral families}
\label{app:familyplots}

Generic integral families needed for the NNLO $q\bar{q}\to t\bar{t}H$:\\
\begin{tabular}{c c c c c c}
    \raisebox{0.5ex}{\scalebox{0.75}{\input{figs/b1.tikz}}} & \raisebox{0.5ex}{\scalebox{0.75}{\input{figs/b3.tikz}}} & \raisebox{0.5ex}{\scalebox{0.75}{\input{figs/b4.tikz}}} & \raisebox{0.5ex}{\scalebox{0.75}{\input{figs/b8.tikz}}} & \raisebox{0.5ex}{\scalebox{0.75}{\input{figs/b12.tikz}}} \\
    {b1} & {b3} & {b4} & {b8} & {b12} \\
    \\
    \raisebox{0.5ex}{\scalebox{0.75}{\input{figs/b13.tikz}}} & \raisebox{0.5ex}{\scalebox{0.75}{\input{figs/b19.tikz}}} & \raisebox{0.5ex}{\scalebox{0.75}{\input{figs/b20.tikz}}} & \raisebox{0.5ex}{\scalebox{0.75}{\input{figs/b25.tikz}}}  & \raisebox{0.5ex}{\scalebox{0.75}{\input{figs/b38.tikz}}}  \\
    {b13} & {b19} & {b20} & {b25} & {b38}  \\
    \\
    \raisebox{0.5ex}{\scalebox{0.75}{\input{figs/b39.tikz}}} & \raisebox{0.5ex}{\scalebox{0.75}{\input{figs/b41.tikz}}} & \raisebox{0.5ex}{\scalebox{0.75}{\input{figs/b45.tikz}}} & \raisebox{0.5ex}{\scalebox{0.75}{\input{figs/b46.tikz}}} & \raisebox{0.5ex}{\scalebox{0.75}{\input{figs/b54.tikz}}} \\
    {b39} & {b41} & {b45} & {b46} & {b54} \\
    \\
    \raisebox{0.5ex}{\scalebox{0.75}{\input{figs/b55.tikz}}} & \raisebox{0.5ex}{\scalebox{0.75}{\input{figs/b58.tikz}}} & \raisebox{0.5ex}{\scalebox{0.75}{\input{figs/b62.tikz}}} & \raisebox{0.5ex}{\scalebox{0.75}{\input{figs/b75.tikz}}} & \raisebox{0.5ex}{\scalebox{0.75}{\input{figs/b78.tikz}}} \\
    {b55} & {b58} & {b62} & {b75} & {b78} \\
    \\
    \raisebox{0.5ex}{\scalebox{0.75}{\input{figs/b79.tikz}}} & \raisebox{0.5ex}{\scalebox{0.75}{\input{figs/b81.tikz}}} & \raisebox{0.5ex}{\scalebox{0.75}{\input{figs/b82.tikz}}} & \raisebox{0.5ex}{\scalebox{0.75}{\input{figs/b84.tikz}}} & \raisebox{0.5ex}{\scalebox{0.75}{\input{figs/b87.tikz}}} \\
    {b79} & {b81} & {b82} & {b84} & {b87} \\
    \\
    \raisebox{0.5ex}{\scalebox{0.75}{\input{figs/b90.tikz}}} & \raisebox{0.5ex}{\scalebox{0.75}{\input{figs/b94.tikz}}} & \raisebox{0.5ex}{\scalebox{0.75}{\input{figs/b98.tikz}}} \\
    {b90} & {b94} & {b98} \\
\end{tabular}

\subsection{Numerical results at example phase-space points}
\label{app:pspoint}
  
In this section we provide results for both the renormalised and bare
amplitudes at three example points.
The first point is a rationalised version of the centre point
from \ref{eq:centre}; it is given by
\begin{equation}
    x_{12} = 778/21, \;
    x_{23} = -119/16, \;
    x_{35} = 232/17, \;
    x_{54} = 111/43, \;
    x_{41} = -184/25,
    \label{eq:example-point-1}
\end{equation}
which corresponds to
\begin{equation}
    \beta^2   \approx 0.79996, \;
    \fstt     \approx 0.69990, \;
    \theta_H  \approx 0.64365, \;
    \theta_t  \approx 0.45105, \;
    \varphi_t \approx 0.79563 \, .
    \label{eq:example-point-1PS}
\end{equation}
The second point is
\begin{equation}
    x_{12} = 509/20, \;
    x_{23} = -187/15, \;
    x_{35} = 156/23, \;
    x_{54} = 97/29, \;
    x_{41} = -300/29;
    \label{eq:example-point-2}
\end{equation}
\begin{equation}
    \beta^2   \approx 0.70880, \;
    \fstt     \approx 0.80662, \;
    \theta_H  \approx 1.68565, \;
    \theta_t  \approx 0.96986, \;
    \varphi_t \approx 1.48048 \, .
\end{equation}
The third point is
\begin{equation}
    x_{12} = 1045/49, \;
    x_{23} = -332/45, \;
    x_{35} = 27/7, \;
    x_{54} = 244/31, \;
    x_{41} = -259/22,
    \label{eq:example-point-3}
\end{equation}
\begin{equation}
    \beta^2   \approx 0.65250, \;
    \fstt     \approx 0.54757, \;
    \theta_H  \approx 0.49515, \;
    \theta_t  \approx -0.53510, \;
    \varphi_t \approx 0.05988 \, .
\end{equation}
For all three points we set $m_H^2/m_t^2 = 12/23$, $m_t^2=1$, and $\mu^2=s_{12}/4$.
The $\eps^0$ parts of the renormalised results are given in~\ref{tab:results_renorm}.
%\footnote{The numbers in the parentheses indicate the uncertainty of the final digits: $1.234(5)$ means $1.234\pm0.005$.}
The corresponding bare results are given in~\ref{tab:results_bare}.

\begin{table}
\centering
\scalebox{0.9}{
\begin{tabular}{llll}
\toprule
\multicolumn{1}{c}{\textsubscript{Amplitude}\textbackslash\textsuperscript{Point}} & \multicolumn{1}{c}{1} & \multicolumn{1}{c}{2} & \multicolumn{1}{c}{3}\tabularnewline
\midrule
\midrule
    ${\cal A}$           & $+6.035834626751872$ & $+5.245314383987918$ & $+4.787293854651411$ \tabularnewline
    ${\cal B}_{l}$       & $-1.12818724680(1)$  & $-0.980427258104(8)$ & $-0.894816410239(8)$ \tabularnewline
    ${\cal B}_{h}$       & $+5.04894231999(3)$  & $+1.89923463609(2)$  & $-0.53804290284(2)$ \tabularnewline
    ${\cal B}_{C_F}$     & $+33.2585182(3)$     & $+27.14075277(9)$    & $+26.3240645(4)$ \tabularnewline
    ${\cal B}_{C_A}$     & $-9.9464334(2)$      & $-10.03564577(5)$    & $-10.4615214(2)$ \tabularnewline
    ${\cal B}_{d_{33}}$  & $-20.3981186(7)$     & $+3.27341269(5)$     & $+6.4743383(4)$ \tabularnewline
    ${\cal C}_{ll}$      & $-26.2652567(5)$     & $-22.8252656(6)$     & $-20.8321652(6)$ \tabularnewline
    ${\cal C}_{lh}$      & $-52.05077(8)$       & $-42.5368(1)$        & $-33.0462(2)$ \tabularnewline
    ${\cal C}_{hh}$      & $-19.97468(2)$       & $-19.18048(3)$       & $-14.51338(2)$ \tabularnewline
    ${\cal C}_{lC_F}$    & $+77.28(4)$          & $+98.12(5)$          & $+56.70(3)$ \tabularnewline
    ${\cal C}_{lC_A}$    & $-29.78(2)$          & $-29.09(1)$          & $-18.77(2)$ \tabularnewline
    ${\cal C}_{ld_{33}}$ & $+62.46(1)$          & $-16.896(4)$         & $-27.186(6)$ \tabularnewline
    ${\cal C}_{hC_F}$    & $+269.333(4)$        & $+191.919(5)$        & $+98.002(6)$ \tabularnewline
    ${\cal C}_{hC_A}$    & $-51.512(8)$         & $-32.936(7)$         & $-10.338(8)$ \tabularnewline
    ${\cal C}_{hd_{33}}$ & $+62.444(4)$         & $-12.869(4)$         & $-15.198(6)$ \tabularnewline
\bottomrule
\end{tabular}
}
\caption{%
Results for the renormalised amplitudes at the three phase-space points given in \ref{app:pspoint}.
The numbers in the parentheses indicate the uncertainty of the final
digits, e.g. $1.234(5)$ means $1.234\pm0.005$.\label{tab:results_renorm}}
\end{table}

\begin{table}
\centering
\scalebox{0.9}{
\begin{tabular}{lllll}
\toprule
\multicolumn{2}{c}{\textsubscript{Amplitude}\textbackslash\textsuperscript{Point}} & \multicolumn{1}{c}{1} & \multicolumn{1}{c}{2} & \multicolumn{1}{c}{3}\tabularnewline
\midrule
\midrule
$A$           & $\eps^{0}$  & $+6.035834626751872$  & $+5.245314383987918$  & $+4.787293854651411$  \tabularnewline
              & $\eps^{1}$  & $-7.519314317124661$  & $-6.966230138192461$  & $-5.852146603415052$  \tabularnewline
$B_{l}$       & $\eps^{-1}$ & $-4.023889751168(3)$  & $-3.496876255992(3)$  & $-3.191529236434(2)$  \tabularnewline
              & $\eps^{0}$  & $+4.979593966290(1)$  & $+3.302193921836(8)$  & $+2.112504306161(7)$  \tabularnewline
              & $\eps^{1}$  & $+16.27790898725(2)$  & $+15.63480325240(1)$  & $+14.56344620751(1)$  \tabularnewline
$B_{h}$       & $\eps^{-1}$ & $-4.023889751168(4)$  & $-3.496876255992(3)$  & $-3.191529236434(3)$  \tabularnewline
              & $\eps^{0}$  & $+2.19990820228(3)$   & $-0.28883877784(1)$   & $-2.87224904596(2)$   \tabularnewline
              & $\eps^{1}$  & $+16.8522808781(2)$   & $+15.04583104135(6)$  & $+10.31556316624(4)$  \tabularnewline
$B_{C_F}$     & $\eps^{-2}$ & $-6.035834626752(4)$  & $-5.245314383988(4)$  & $-4.787293854651(3)$  \tabularnewline
              & $\eps^{-1}$ & $+47.50645250594(3)$  & $+36.76901095667(3)$  & $+28.96795514643(2)$  \tabularnewline
              & $\eps^{0}$  & $+82.8900101(3)$      & $+71.98685363(9)$     & $+69.3832423(4)$      \tabularnewline
              & $\eps^{1}$  & $+10.1664278(5)$      & $-12.2056345(6)$      & $+16.743047(1)$       \tabularnewline
$B_{C_A}$     & $\eps^{-1}$ & $+6.38676000270(2)$   & $+6.29876756710(1)$   & $+5.90321310713(1)$   \tabularnewline
              & $\eps^{0}$  & $-25.4696877(2)$      & $-22.15585574(5)$     & $-19.5366653(2)$      \tabularnewline
              & $\eps^{1}$  & $+41.3310335(7)$      & $+33.4761392(3)$      & $+24.472308(1)$       \tabularnewline
$B_{d_{33}}$  & $\eps^{-1}$ & $+12.49071290385(3)$  & $-2.49052049845(2)$   & $-4.40305429926(2)$   \tabularnewline
              & $\eps^{0}$  & $-23.7968559(7)$      & $+3.01592475(5)$      & $+5.2408179(4)$       \tabularnewline
              & $\eps^{1}$  & $+8.456396(3)$        & $+3.9717895(3)$       & $+4.764632(2)$        \tabularnewline
$C_{ll}$      & $\eps^{-2}$ & $+2.68259318(2)$      & $+2.33125085(2)$      & $+2.12768616(1)$      \tabularnewline
              & $\eps^{-1}$ & $-3.29754111(7)$      & $-1.30682292(6)$      & $-0.21571834(6)$      \tabularnewline
              & $\eps^{0}$  & $-48.1798268(5)$      & $-43.5116000(6)$      & $-39.7488578(6)$      \tabularnewline
$C_{lh}$      & $\eps^{-2}$ & $+5.36518633489(1)$   & $+4.662501674656(4)$  & $+4.255372315246(9)$  \tabularnewline
              & $\eps^{-1}$ & $-2.88884(2)$         & $+2.17440(2)$         & $+6.21486(4)$         \tabularnewline
              & $\eps^{0}$  & $-94.30589(8)$        & $-80.2102(1)$         & $-61.3569(2)$         \tabularnewline
$C_{hh}$      & $\eps^{-2}$ & $+2.682593167445276$  & $+2.3312508373279632$ & $+2.127686157623(2)$  \tabularnewline
              & $\eps^{-1}$ & $+0.4087065(2)$       & $+3.4812207(5)$       & $+6.4306195(5)$       \tabularnewline
              & $\eps^{0}$  & $-45.35684(2)$        & $-35.29000(3)$        & $-18.74365(2)$        \tabularnewline
$C_{lC_F}$    & $\eps^{-3}$ & $+5.02986221(5)$      & $+4.37109534(4)$      & $+3.98941156(4)$      \tabularnewline
              & $\eps^{-2}$ & $-44.5433305(3)$      & $-32.8089814(2)$      & $-24.8738667(2)$      \tabularnewline
              & $\eps^{-1}$ & $-242.637(6)$         & $-216.388(7)$         & $-200.168(5)$         \tabularnewline
              & $\eps^{0}$  & $-21.44(4)$           & $-49.71(5)$           & $-154.50(3)$          \tabularnewline
$C_{lC_A}$    & $\eps^{-2}$ & $-10.075326(1)$       & $-9.5042367(7)$       & $-8.8287811(7)$       \tabularnewline
              & $\eps^{-1}$ & $+73.197(3)$          & $+59.106(1)$          & $+51.344(2)$          \tabularnewline
              & $\eps^{0}$  & $-43.93(2)$           & $-2.58(1)$            & $+26.79(2)$           \tabularnewline
$C_{ld_{33}}$ & $\eps^{-2}$ & $-12.490710(4)$       & $+2.4905207(7)$       & $+4.403055(1)$        \tabularnewline
              & $\eps^{-1}$ & $+19.527(3)$          & $-0.896(1)$           & $-1.104(2)$           \tabularnewline
              & $\eps^{0}$  & $+50.43(1)$           & $-21.997(4)$          & $-32.660(6)$          \tabularnewline
$C_{hC_F}$    & $\eps^{-3}$ & $+8.04777950232(2)$   & $+6.993752511979(8)$  & $+6.38305847286(2)$   \tabularnewline
              & $\eps^{-2}$ & $-46.649196(7)$       & $-32.454596(5)$       & $-21.50151(2)$        \tabularnewline
              & $\eps^{-1}$ & $-271.8951(7)$        & $-243.9427(3)$        & $-222.9754(3)$        \tabularnewline
              & $\eps^{0}$  & $+58.513(4)$          & $-59.977(5)$          & $-249.499(6)$         \tabularnewline
$C_{hC_A}$    & $\eps^{-2}$ & $-8.51570(3)$         & $-8.398366(6)$        & $-7.87093(3)$         \tabularnewline
              & $\eps^{-1}$ & $+65.1827(5)$         & $+46.6892(3)$         & $+31.8517(9)$         \tabularnewline
              & $\eps^{0}$  & $-29.121(8)$          & $+14.636(7)$          & $+36.464(9)$          \tabularnewline
$C_{hd_{33}}$ & $\eps^{-2}$ & $-16.6542837(3)$      & $+3.320693935(9)$     & $+5.87073910(8)$      \tabularnewline
              & $\eps^{-1}$ & $+9.6379(3)$          & $+1.5642(3)$          & $+4.9777(3)$          \tabularnewline
              & $\eps^{0}$  & $+61.991(4)$          & $-17.139(4)$          & $-18.893(6)$          \tabularnewline
\bottomrule
\end{tabular}
}
\caption{Results for the bare amplitudes at the example points from \ref{app:pspoint}.\label{tab:results_bare}}
\end{table}
\clearpage